\documentclass[11pt]{article}
\usepackage{jheppub}

\usepackage{amssymb}
\usepackage{amsmath}
\usepackage{amsthm}
\usepackage{mathtools}
\usepackage[usenames,dvipsnames]{xcolor}
\usepackage{tikz}
\usepackage{tikz-cd}
\usetikzlibrary{shapes.geometric, arrows}
\usepackage{appendix}
\usepackage[dvipsnames]{xcolor} 
\usepackage[utf8]{inputenc}
\usepackage{dsfont}
\usepackage{cancel}
\usepackage{comment}
\usepackage{graphicx}
\usepackage{graphbox}
\usepackage{lscape}

\usepackage{hyperref}
\hypersetup{
	colorlinks = True,
   	citecolor = magenta,
   	linkcolor = cyan,
	urlcolor = cyan,
	bookmarks=true
}

\newcommand{\nn}{\nonumber}
\renewcommand{\d}{\text{d}}
\newcommand{\D}{\text{D}}
\newcommand{\res}{\text{Res}}
\newcommand{\vep}{\varepsilon}
\newcommand{\vphi}{\varphi}
\newcommand{\vth}{\vartheta}
\newcommand{\la}{\langle}
\newcommand{\ra}{\rangle}
\newcommand{\mbf}[1]{\mathbf{#1}}
\newcommand{\bs}[1]{\boldsymbol{#1}} 
\newcommand{\mat}[1]{\underline{\boldsymbol{#1}}}
\newcommand{\del}[1]{\nabla}
\renewcommand{\c}[1]{\check{#1}}
\newcommand{\cdel}[1]{{\check{\nabla}}}
\newcommand{\com}{\check{\omega}}
\newcommand{\cu}{\check{u}}
\newcommand{\cvphi}{\check{\vphi}}
\newcommand{\cphi}{\check{\phi}}
\renewcommand{\vec}[1]{\mbf{#1}}

\newcommand{\ibp}{\mathrm{IBP}}

\def\ap#1{{\color{violet}[AP: #1]}}
\def\sd#1{{\color{blue}[SD: #1]}}

\title{Cosmology meets cohomology}

\author{Shounak De}
\emailAdd{shounak\_de@brown.edu}
\author{and Andrzej Pokraka}
\emailAdd{andrzej\_pokraka@brown.edu}

\affiliation{Department of Physics, Brown University, \\
	182 Hope Street, Providence, RI 02912, U.S.A.}

\abstract{The cosmological polytope and bootstrap programs have revealed interesting connections between positive geometries, modern on-shell methods and bootstrap principles studied in the amplitudes community with the  wavefunction of the Universe in toy models of FRW cosmologies.
To compute these FRW correlators, one often faces integrals that are too difficult to evaluate by direct integration. Borrowing from the Feynman integral community, the method of (canonical) differential equations provides an efficient alternative for evaluating these integrals. Moreover, we further develop our geometric understanding of these integrals by describing the associated \emph{relative} twisted cohomology. Leveraging recent progress in our understanding of relative twisted cohomology in the Feynman integral community, we give an algorithm to predict the basis size and simplify the computation of the differential equations satisfied by FRW correlators. 
}

\begin{document}

\maketitle

\section{Introduction}

Over the past decade, an increasing number of concepts from the theory of hypergoemetric functions and more broadly algebraic geometry have entered modern methods for studying scattering amplitudes. In particular, the framework of twisted cohomology (originally formulated to study the linear relations among hypergoemetric functions) has been used to construct CHY amplitudes \cite{Cachazo:2013gna, Mizera:2017rqa}, provide a geometric framework for understanding KLT relations in string amplitudes \cite{Kawai:1985xq, Mizera:2017cqs, Casali:2019ihm, Stieberger:2022lss}, the double copy \cite{Mizera:2019blq} and more.

In the Feynman integral community, twisted intersection theory 
\cite{Mastrolia:2018uzb, Frellesvig:2019kgj, Mizera:2019gea, Mizera:2019vvs, Frellesvig:2020qot, Weinzierl:2020xyy, Chestnov:2022xsy, Chen2022}
provides an algebraic alternative to integral reduction\footnote{Integral reduction is the process of decomposing an arbitrary integral into a linear combination of basis (also called master) integrals that only need to be computed once.}
via integration-by-parts (IBP) identities 
\cite{Chetyrkin:1981qh, Gehrmann:1999as, Laporta:2000dsw}. 
While these techniques are still being optimized for applications to physics, they have the potential to circumvent solving large linear systems of IBPs thereby streamlining one of the most expensive steps in the pipeline for multi-loop phenomenological calculations. 

Recently, the procedure for calculating intersection numbers of Feynman integrals has been streamlined by using \emph{relative} twisted cohomology \cite{Caron-Huot:2021xqj, Caron-Huot:2021iev}. By working with relative twisted cohomology, one can set the regulator in previous works to zero at the start of the calculation. As a consequence, many intermediate quantities vanish or simplify, streamlining the overall algorithm. 

The last decade has also seen the discovery of rich geometrical structures in the study of flat-space scattering amplitudes which provide answers to elegant questions posed purely in the kinematic space. While such insights have transformed our understanding of the flat-space S-matrix, the situation for cosmological correlators is still underdeveloped. The lack of such methods for cosmology and the complexity of even tree-level results\footnote{One already encounters branch cuts at the level of trees in cosmological correlators whereas their flat-space counterparts are rational functions in the kinematic invariants.} means that the repository of theoretical data is significantly smaller. 
The problem is further complicated by the fact that inflationary correlators live on the late-time boundary of de Sitter, which is space-like. The understanding of the holographic dual to theories with space-like boundaries is still in its infancy, while the boundary description for anti-de Sitter space is well established due to the time-like nature of its boundary.
Having said that, the cosmological polytope and bootstrap programs have taken important steps towards a boundary description of inflationary correlators.

While de Sitter models the accelerated expansion of the early Universe, the Friedmann-Robertson–Walker (FRW)\footnote{While the most commonly used term is the FRW metric, it is also sometimes called the FLRW metric after Friedmann, Lema\^{i}tre, Robertson and Walker.} metric describes a homogeneous, isotropic, expanding (or contracting) Universe. The FRW spacetime is often considered to be the standard model of cosmology post inflation.
Field theoretic correlation functions in FRW cosmologies represent quantum fluctuations, which in the early Universe are thought to have sourced the observed distribution of matter. 
Comparing these predictions to current and future cosmological datasets will help constrain new physics beyond the Standard Model.

Unlike the de Sitter geometry (which has provided a rich playground for studying inflationary correlators in the literature so far), FRW Universes are not maximally symmetric spacetimes and have a reduced isometry group. This reduces the applicability of the cosmological bootstrap program to derive differential equations for correlators beyond de Sitter space. However, it turns out that correlators in FRW cosmologies (the main focus of this work\footnote{We will consider FRW cosmologies with a future spacelike boundary where our boundary observables live.}) furnish some of the simplest examples of relative twisted cohomology.\footnote{Mathematically, FRW correlators are integrals associated to hyperplane arrangements where the coordinate hyperplanes are twisted and all other divisors are left untwisted.}
Thus, they provide fertile testing ground for developing our understanding of relative twisted cohomologies, which in turn will deepen our understanding of the underlying physics. 

While the standard tool for computing cosmological correlation functions is perturbation theory, the past decade has seen novel computational methods with interesting and fruitful connections to scattering amplitudes (e.g., symbology \cite{Hillman:2019wgh}) and mathematics (the combinatorial description in terms of the cosmological polytope \cite{ArkaniHamed2017,Arkani-Hamed:2018bjr,Benincasa:2019vqr,Benincasa:2021qcb}). 
Many of these methods manifest the simplicity of the final answer that is hidden at intermediate steps in perturbation theory. 
In particular, the singularity structure of these correlators is particularly evident in the polytopal picture and encodes principles of unitary (by cutting rules) as well as causality (Steinmann relations). 

Our main goal is two fold. First, we aim to provide efficient methods for computing the differential equations of FRW wavefunction coefficients.
Deriving differential equations for cosmological correlators was first considered in the upcoming work \cite{nimafriends} (also see the talks \cite{ptalk, blectures, ltalk}). 
We also hope that by formally constructing the cohomology associated to the integrals appearing in the FRW wavefunction coefficients and elucidating their mathematical structure will lead to a deeper physical understanding of these objects. 
Secondly, our techniques are quite general and are applicable to many integrals seen in both physics and mathematics. Thus, we expect this work to be useful beyond the current setting of cosmological correlators.

While primarily aimed at physicists, we think that much of this work will interest mathematicians. 
While many of our claims have been well tested, this paper lacks formal proofs and it would be interesting to understand these claims more formally. 
As far as the authors know, \cite{Matsumoto:aa} is the only paper discussing relative twisted cohomology in the mathematics literature.  

The paper is organized as follows. In section \ref{sec:CosmoWavefunction}, we set up the toy model of conformally coupled scalars in a general FRW background studied in this work and explain how to uplift flat-space wavefunction coefficients to their FRW counterparts. 
In section \ref{sec:4pt}, we study the integral representation of the two-site/four-point FRW integral in detail. To evaluate this integral, we derive the differential equation satisfied by a basis of integrals for the corresponding integral family. We obtain these differential equations using three different perspectives: twisted cohomology and standard IBPs in section \ref{sec:4ptBad}, the dual (relative) twisted cohomology and dual IBPs in section \ref{sec:4ptSimp}, and intersection numbers in section \ref{sec:4ptNoIBP}. The dual perspective (section \ref{sec:4ptSimp}) yields computational advantages as well as novel geometric understanding and is central to this work. 
In section \ref{sec:4ptIntegration}, we explicitly integrate these differential equations to obtain the first correction to the de Sitter two-site/four-point function.
In section \ref{sec:further examples}, we provide two more worked examples: the one-loop two-site/two-point and the tree-level three-site/five-point FRW integrals. In this section, we illustrate how to use the techniques developed in section \ref{sec:4pt} on more complicated examples. 
We conclude in section \ref{sec:conclusion} with a discussion of our results and directions for future work. 

\section{Cosmological correlators and the Wavefunction of the Universe \label{sec:CosmoWavefunction}}

We consider the model of a conformally-coupled scalar field in a general FRW cosmology with non-conformal polynomial interactions, studied extensively in \cite{ArkaniHamed2017,Arkani-Hamed:2018bjr,Hillman:2019wgh,Benincasa:2018ssx}. The action for such a theory in a $(d+1)$-dimensional spacetime is 
\begin{align}
    \mathcal{S} = \int \d^d x \, \d\eta \sqrt{-g} \left[-\frac{1}{2} g^{\mu \nu} \partial_{\mu} \phi \partial_{\nu} \phi - \frac{1}{2} \xi R \phi^2 - \sum_{k \geq 3} \frac{\lambda_k}{k!} \phi^k \right]~,
    \label{eq:actioninFRW}
\end{align}
with the FRW metric and the constant $\xi$ defined as
\begin{align}
    \d s^2 \equiv g_{\mu \nu} \d x^{\mu} \d x^{\nu} = a^2(\eta) \left[-\d \eta^2 + \d x_i \d x^i\right], \quad \xi = \frac{d-1}{4 d}~.
    \label{eq:FRWmetricsI}
\end{align}
The FRW metric above has been written in comoving coordinates with conformal time $\eta \in (-\infty, 0]$ and the index $i=1,\dots,d$ runs over the spatial dimensions. $a(\eta)$ denotes the scale factor and the choice of its representation in various stages of the Universe's evolution will be crucial for the class of integrals dealt with in our paper. Importantly, the choice of $\xi$ sets the scalar to be conformally-coupled thereby making the action $\mathcal{S}$ conformally equivalent to the following flat-space action
\begin{align}
    S[\phi] = \int \d^d x \, \d\eta \left[-\frac{1}{2} (\partial \phi)^2 -\sum_{k \geq 3} \frac{\lambda_k (\eta)}{k!} \phi^k\right]~.
    \label{eq:actioninflatspace}
\end{align}
More precisely, the conformal transformation 
\begin{align}
    g_{\mu \nu} \rightarrow a^2(\eta) g_{\mu \nu} \ ,\quad \phi \rightarrow a^{- \Delta}(\eta) \phi \ ,\quad \Delta = \frac{d-1}{2}~,
\end{align}
allows one to rewrite the original action $\mathcal{S}$ (\ref{eq:actioninFRW}) as the action of a massless scalar field in $(d+1)$-dimensional flat-space (\ref{eq:actioninflatspace}) with time-dependent couplings given by
\begin{align}
    \lambda_k(\eta) \equiv \lambda_k [a(\eta)]^{(2-k)\Delta+2}~.
    \label{eq:timedepncouplings}
\end{align}
It is worth stressing that the conformal equivalence between the original action $\mathcal{S}$ (\ref{eq:actioninFRW}) and the flat-space action $S[\phi]$ (\ref{eq:actioninflatspace}) allows us to do meaningful computations in a general FRW cosmology (described by the scale factor $a(\eta)$) simply by doing flat-space perturbation theory (albeit with time-dependent couplings given by (\ref{eq:timedepncouplings})), as we shall see below. 

Of central importance, is the wavefunction of the universe $\Psi[\Phi]$. Formally, this is computed as a path integral by integrating over all bulk field configurations $\phi(\vec{x},\eta)$ with non-vanishing Dirichlet boundary condition in the future $\phi(\vec{x},\eta=0)=\Phi(\vec{x})$. It is standard to work in momentum space where the wavefunction can be expanded as
\begin{align}
    \Psi[\Phi] = \int_{\phi(-\infty(1-i\epsilon))}^{\phi(0)=\Phi} \mathcal{D} \phi \, e^{i S[\phi]} \equiv \exp{\left[i \sum_n \frac{1}{n!} \int \prod_{i} \d^d \vec{k}_i \, \Phi(\vec{k}_i) \, \tilde{\psi}_n(\{\vec{k}_i\})\right]}~,
    \label{eq:Uniwavefunc}
\end{align}
and the standard $i \epsilon$ prescription selects the adiabatic/Bunch-Davies/Hartle-Hawking vacuum at the early-time boundary $\eta \to -\infty$. 
The kernels $\tilde{\psi}_n$ are called \textit{wavefunction coefficients} and $\{\vec{k}_i\}$ denotes the set of all \textit{spatial} momentum. 
Due to \textit{spatial} translation invariance on the future boundary, the wavefunction coefficients contain an overall momentum-conserving $\delta$-function. It is useful to extract this $\delta$-function and compute the ``stripped'' wavefunction coefficients $\psi_n$:
    \begin{align}
        \tilde{\psi}_n(\{\vec{k}_i\}) = \delta^{(d)}\left(\sum_{i=1}^n \vec{k}^{(i)}\right) {\psi}_n(\{\vec{k}_i\})~.
    \end{align}
Our main objective is the computation of these stripped wavefunction coefficients. 

Traditionally, the wavefunction coefficients $\psi_n$ are obtained perturbatively, using Feynman diagrammatics, which we review here. 
We approximate the path integral (\ref{eq:Uniwavefunc}) by its saddle point $\Psi[\Phi] \approx \exp{(iS[\Phi_{\textrm{cl}}])}$ where $\Phi_{\textrm{cl}}$ is the boundary profile corresponding to the classical solution to the equations of motion $\phi_{\textrm{cl}}$.
To derive the Feynman rules, we write the classical solution as 
 \begin{align}
        \phi_{\textrm{cl}}(\vec{k},\eta) = \mathcal{K}(E,\eta) \Phi(\vec{k}) 
        + \int \d\eta^{\prime} \mathcal{G}(E;\eta,\eta^{\prime}) \frac{\delta S_{\textrm{int}}}{\delta \phi(\vec{k},\eta^{\prime})}~. 
        \label{eq:classicalsolution}
    \end{align}
Substituting the above solution into the action, one can verify that the path integral evaluates to the expression in (\ref{eq:Uniwavefunc}) by taking functional derivatives of the path integral with respect to the boundary value $\Phi(\vec{k})$ (which acts like a source term). To ensure that the solution (\ref{eq:classicalsolution}) satisfies the correct Dirichlet boundary conditions, one needs to appropriately solve for the two distinct Green's functions, the bulk-to-boundary propagator $\mathcal{K}(E_v,\eta_v)$ associated with a vertex $v$ and the bulk-to-bulk propagator $\mathcal{G}(E_e,\eta_v,\eta^{\prime}_v)$ associated with an edge $e$ connecting a vertex at time $\eta_v$ to another one at time $\eta^{\prime}_v$. Explicitly, they are given by the expressions
    \begin{align}
       \mathcal{K}(E_v,\eta_v) &= e^{iE_v \eta_v}~, \label{eq:bulktoboundaryprop} \\
       \mathcal{G}(E_e,\eta_v,\eta_v^{\prime}) &= \frac{1}{2E_e} \left(e^{{-}iE_e(\eta_v{-}\eta_v^{\prime})} \theta\left(\eta_v {-} \eta_v^{\prime}\right) + e^{{-}iE_e(\eta_v^{\prime}{-}\eta_v)} \theta\left(\eta_v^{\prime}{-}\eta_v\right) -e^{iE_e(\eta_v{+}\eta_v^{\prime})}\right)~,
       \label{eq:bulktobulkprop}
    \end{align}
where $E_v=\sum_{j}|\vec{k}_j|$ denotes the sum of the energies of $j$ external lines connected to the vertex $v$, and $E_e=|\vec{k}_e|$ is the energy of an internal line. The bulk-to-boundary propagator, $\mathcal{K}(E,\eta)$, solves the linearized, homogeneous EOM, oscillates with positive frequency in the far past, and approaches unity at the late time boundary as $\eta \to 0$. The bulk-to-bulk propagator, $\mathcal{G}(E,\eta,\eta^{\prime})$, solves the inhomogeneous wave equation $\sqrt{-g}(\Box - m^2)\mathcal{G}(E,\eta,\eta^{\prime}) = i \delta(\eta-\eta^{\prime})$, subject to the boundary condition that it vanishes identically when either of its time arguments approach zero. The combined boundary conditions on the Green's functions guarantee that the path integral in (\ref{eq:Uniwavefunc}) as well as the classical solution (\ref{eq:classicalsolution})   satisfy the correct Dirichlet boundary conditions. 
The problem of computing the wavefunction coefficients $\psi_n$ is now thus reduced to an exercise in perturbative diagrammatic expansion using Feynman rules for the fluctuations of the theory at hand. 
We note that to compute these wavefunction contributions, it suffices to implement flat-space perturbation theory which thereby surpasses the need to introduce complicated Green's functions for a general FRW spacetime. As we shall see in the subsequent sections, for various FRW cosmologies this is facilitated by an operation involving integration over the external energies that accounts for the time-dependent couplings (\ref{eq:timedepncouplings}) of the theory.  


\subsection{FRW cosmologies \label{sec:FRWintro}}
In this work, we study the analytic structure of integrals associated to the wavefunction coefficients in a general FRW Universe, going beyond the inflationary expansion modelled by a de Sitter Universe. 
Before presenting the integrals tackled in this paper, we give a brief description of the FRW backgrounds describing the Universe's evolution at its various stages: 
\paragraph{de Sitter Universe} de Sitter (dS) space models the accelerated expansion of the inflationary Universe to a good approximation and thus provides a starting point for a boundary description of inflationary correlators. The isometry group of dS space has been utilized to exploit the conformal symmetry of cosmological correlators in the recent cosmological bootstrap program \cite{Arkani-Hamed:2018kmz,Baumann:2022jpr,Pajer:2020wnj}. We recall that the de Sitter line element in flat slicing is given by
\begin{align}
    \d s^2 = \frac{-\d \eta^2 + \d x_i \d x^i}{\eta^2}~,
    \label{eq:deSittermetric}
\end{align}
with $\eta \in (-\infty,0]$ and the Hubble scale set to unity. Comparing with the FRW metric (\ref{eq:FRWmetricsI}), one sees that the scale factor behaves as $a(\eta) = 1/\eta$ for a de Sitter Universe. Much of the theoretical data concerning cosmological correlators come from our understanding of polynomial interactions in dS space. A particularly well-studied model is the $\lambda_3 \phi^3$ theory in dS$_4$, which provides a rich framework for understanding the analytic structure of inflationary correlators \cite{ArkaniHamed2015}.

\paragraph{Radiation/Matter dominated Universe} For most of its history the Universe was dominated by a single component, as suggested by the different scalings of the energy density $\rho$ with the scale factor $a(\eta)$. A phase of radiation domination (RD) where $\rho \propto a^{-4}$ was followed by an era of matter domination (MD) where $\rho \propto a^{-3}$. The dependencies of the scale factor on the conformal time in these phases are obtained by integrating the first Friedmann equation for a flat, single-component Universe \cite{Baumann:2022mni}. This leads to the following FRW solutions for the scale factor
\begin{align}
    a(\eta) \propto
    \begin{cases}
        \eta \quad (\textrm{RD}) \\
        \eta^2 \quad (\textrm{MD})~,
        \label{eq:RD,MDscalefactors}
    \end{cases}
\end{align}
where the proportionality factors depend on dimensionless density parameters and will be set to unity for our discussions. 

Given the above dependencies of the scale factor $a(\eta)$, the different FRW solutions (\ref{eq:FRWmetricsI}) corresponding to a de Sitter (\ref{eq:deSittermetric}), flat, radiation or a matter-dominated Universe (\ref{eq:RD,MDscalefactors}) can be encapsulated by the following metric parameterized by $\vep$
\begin{align}
    \d s^2 = \eta^{-2 \vep} \left[\frac{-\d\eta^2+\d x_i \d x^i}{\eta^2}\right] 
    \begin{cases}
        \vep = 0 \quad (\textrm{dS}) \\
        \vep = -1 \quad (\textrm{flat})  \\
        \vep = -2 \quad (\textrm{RD})  \\
        \vep = -3 \quad (\textrm{MD})~,
    \end{cases}
    \label{eq:FRWmetricsII}
\end{align}
such that the scale factor is 
\begin{align} 
    a(\eta) = \frac{1}{\eta^{1+\vep}} \, .
\end{align}
While the differential equations (DEQ) that we derive in our paper hold for any value of $\vep$, the solution to them is always presented as a Laurent series expansion in $\vep$, as we shall soon see. When solving the DEQ, our analysis will thereby be restricted to cosmologies closely resembling a dS Universe. The consequence of $\vep \sim 0$ is two fold: i) in the analysis of inflationary correlators so far, the literature assumes a perfect dS expansion ($\vep = 0$). However, inflationary cosmology dictates that for inflation to end one has to consider small departures from dS, often leading to inflation being referrred as a quasi-dS period. ii) Even if one models inflation as a perfect dS expansion, computing wavefunction coefficients as a series in $\vep$ could potentially shed crucial insights into physics beyond inflation, i.e., after the reheating surface.

Mathematically, the parameter $\vep$ corresponds to the exponent of the ``\textit{twist}'' associated to the wavefunction coefficients of an FRW Universe.
This identification, allows us to import the technology of \textit{relative twisted cohomology} and \textit{intersection theory} to cosmological correlators. 
Such ideas were first introduced in physics in the context of Feynman integrals. 
There, the parameter $\vep$ is the dimensional regularization parameter.
In both cases, this exponent is generally non-integer giving rise to branch cut singularities on the integration manifold that partially dictates the singularity structure of the final object.

For the sake of simplicity, we shall focus on a theory of conformally coupled scalars described by a cubic $\lambda_3 \phi^3$ interaction in a (1+3)-dimensional FRW Universe described by the action
\begin{align}
\mathcal{S} = \int \d^3 x \, \d\eta \sqrt{-g} \left[-\frac{1}{2} g^{\mu \nu} \partial_{\mu} \phi \partial_{\nu} \phi - \frac{1}{12} R \phi^2 - \frac{\lambda_3}{3!} \phi^3 \right]~.
\label{eq:actioninFRWphi^3theory}
\end{align}
However, it is straightforward to generalize our analysis to theories with polynomial interactions $\lambda_k \phi^k$ in an FRW Universe (\ref{eq:FRWmetricsII}) obeying the condition
\begin{align}
    [a(\eta)]^{(2-k)\frac{(d-1)}{2}+2} = \frac{1}{\eta^{1+\vep}}~,
    \label{eq:scalefactorI}
\end{align}
such that the time-dependent couplings (\ref{eq:timedepncouplings}) are given by $\lambda_k(\eta) = \frac{\lambda_k}{\eta^{1+\vep}}$. 
These time-dependent couplings are most conveniently treated in (temporal) Fourier space
\begin{align}
    \frac{1}{\eta^{1+\vep}} = \frac{(-i)^{1+\vep}}{\Gamma(1+\vep)} \int_{0}^{\infty(1-i \epsilon)} \d x \, x^{\vep} e^{i x \eta}~.
    \label{eq:scalefactorFourier}
\end{align}
Before proceeding to furnish the integral representations for FRW correlators, a few comments are in order regarding the above Fourier-space expression. First, it is worth noting the difference between the FRW parameter $\vep$ and the standard $i \epsilon$ regulator ensuring the convergence of the integral. Secondly, it is crucial to note that the integral converges only for $\vep > -1$ due to the presence of the gamma functions. This is seemingly in contradiction with our boundary choices of $\vep$ denoting the various FRW solutions in (\ref{eq:FRWmetricsII}). However, as we will see in section \ref{sec:4pt}, the differential equations satisfied by the FRW cosmological correlators can be solved for any value of the parameter $\vep$.
In what follows, we suppress the overall constants arising from the gamma functions and the factors of $i$ appearing in (\ref{eq:scalefactorFourier}) to avoid clutter. Given a choice of $\vep$ for the FRW Universe one lives in, such overall factors and coupling constants can be reinstated as they are needed.


\subsection{Cosmological correlators in FRW backgrounds \label{sec:FRWcorr}}

Equipped with the above conceptual preliminaries, we now turn our attention to the construction of the wavefunction coefficients in flat-space and their uplifts to a general FRW cosmology. 

We first define the flat-space $n$-site wavefunction coefficient at $L$-loops $\psi_{n,\textrm{flat}}^{(L)}$ which essentially makes up the \textit{energy integrand} of the FRW wavefunction coefficient
\begin{align}
    \psi_{n,\textrm{flat}}^{(L)}(X_v,Y_e) = \int_{-\infty}^0 \prod_{v \in \mathcal{V}} \d \eta_v \, e^{i X_v \eta_v} \prod_{e \in \mathcal{E}} \mathcal{G}(Y_e, \eta_v, \eta^{\prime}_v)~,
    \label{eq:energyintegrand}
\end{align}
where $\mathcal{V}, \mathcal{E}$  denotes the set of vertices and edges of a given $n$-site, $L$-loop Feynman graph. We also introduce the notation $X_v$ associated with the sum of energies flowing to the boundary from a vertex $v$ and $Y_e$ denote the energies of the edges (internal lines) $e$.

The computation of the nested time integrals given by the formal expression (\ref{eq:energyintegrand}) has been carried out in the literature using several approaches: i) explicitly computing the $3^e$ \textit{bulk} time integrals arising from the three-term structure of the propagator (\ref{eq:bulktobulkprop}), ii) a combinatorial representation to implement \textit{boundary} old-fashioned perturbation theory which one would obtain via the Lippmann-Schwinger kernel \cite{ArkaniHamed2017}, iii) \textit{triangulations} of the cosmological polytope (and its dual) \cite{ArkaniHamed2017,Benincasa:2019vqr,Arkani-Hamed:2018bjr,Juhnke-Kubitzke:2023nrj}. Using any of the approaches, one can compute the two- and three-site flat-space wavefunction coefficients at tree-level to be
\begin{align}
    \psi_{2,\textrm{flat}}^{(0)} 
    &= \scalebox{1.15}{$
    \frac{1}{(X_1{+}X_2) (X_1{+}Y) (X_2{+}Y)}
    $}~, 
    \label{eq:flatspace2-site} 
    \\
    \psi_{3,\textrm{flat}}^{(0)} 
    &= \scalebox{1.25}{$
    \frac{X_1{+}Y_1{+}2X_2{+}Y_2{+}X_3}{(X_1{+}X_2{+}X_3) (X_1{+}Y_1) (X_3{+}Y_2) (X_2{+}Y_1{+}Y_2) (X_1{+}X_2{+}Y_2) (X_2{+}X_3{+}Y_1)}$
    },
    \label{eq:flatspace3-site}
\end{align}
while the two-site contribution at one loop is given by
\begin{align}
    \psi_{2,\text{flat}}^{(1)}
    &= \scalebox{1.25}{$
    \frac{2 \left(X_1+X_2+Y_1+Y_2\right)}{\left(X_1+X_2\right) \left(X_1+X_2+2 Y_1\right) \left(X_1+Y_1+Y_2\right) \left(X_2+Y_1+Y_2\right) \left(X_1+X_2+2 Y_2\right)}$
    }.
    \label{eq:flatspace2-site1-loop}
\end{align}

To compute wavefunction coefficients relevant to FRW cosmologies, we perform an uplift by integrating the \textit{integrand} (\ref{eq:energyintegrand}) over the energies related to the time-dependent coupling constants (\ref{eq:scalefactorFourier}). To account for this extra time-dependence, we denote the \textit{shifted} vertex energies as $E_v = x_v + X_v$ while also tacking on the twist factor $x_v^{\vep}$ at each vertex. As a result, the full FRW wavefunction contributions take the form
\begin{align}
    \psi_{n,\textrm{FRW}}^{(L)}(X_v,Y_e) = \int_0^{\infty} \prod_{v \in \mathcal{V}} \d x_v \, x_v^{\vep} \, \psi_{n,\textrm{flat}}^{(L)}(x_v + X_v,Y_e) = \int_{\mathcal{R}} \, \prod_{v \in \mathcal{V}} \d x_{v} \, x_v^{\vep} \, \psi_{n,\textrm{flat}}^{(L)}(E_v,Y_e)~,
    \label{eq:FRWwavefunccoeff}
\end{align}
where $\mathcal{R}$ defines the rectangular region $E_v > X_v$ for each vertex $v$. 
In the mathematics literature, the integrals \eqref{eq:FRWwavefunccoeff} belong to the class of functions which are often called generalized Euler integrals or GKZ\footnote{Gelfand, Kapranov and Zelevinsky.} hypergeometric functions \cite{Matsubara-Heo:2023ylc}. As we will see in section \ref{sec:4ptIntegration}, the FRW wavefunction coefficients belong to the subset of this function space whose series expansion in $\vep$ involves only multiple polylogarithms.
It is worth noting that \eqref{eq:FRWwavefunccoeff} is actually the spatial loop momentum integrand for loop level corrections. While we will only study the integrals given by \eqref{eq:FRWwavefunccoeff} in this work, the mathematical framework developed here can straightforwardly accommodate the integration over the spatial loop momentum and is left for future work.


As an example of this prescription, consider the uplift of $\psi^{(0)}_{2,\text{flat}} \to \psi^{(0)}_{2,\text{FRW}}$.
Using the two-site flat-space wavefunction coefficient (\ref{eq:flatspace2-site}) and the uplift relation (\ref{eq:FRWwavefunccoeff}), one obtains the explicit realization of the corresponding FRW wavefunction coefficient given by the integral expression 
\begin{align}
    \psi_{2,\text{FRW}}^{(0)}(X_1,X_2;Y) = \int_0^{\infty} \d x_1 \wedge \d x_2 \ \frac{(x_1 x_2)^{\vep}}{(x_1{+}x_2{+}X_1{+}X_2) (x_1{+}X_1{+}Y) (x_2{+}X_2{+}Y)}~.
    \label{eq:2-siteFRW}
\end{align}
The $\vep \to 0$ limit of such integrals reduces to the well-studied cases arising in inflationary correlators in dS$_4$ \cite{Arkani-Hamed:2018kmz,ArkaniHamed2015}. 
It is worth mentioning that while this prescription for the uplift of $\psi_{n,\textrm{flat}} \to \psi_{n,\textrm{FRW}}$ is technically true only for $\vep\neq-1$, the flat-space wavefunction coefficient appears as the coefficient of the leading log divergence in the $\vep\to -1$ limit. Explicitly setting $\vep\to-1+\mu$ in (\ref{eq:2-siteFRW}) yields 
\begin{align}
    \psi^{(0)}_{2,\text{FRW}}(X_1,X_2,;Y)\big|_{\vep \to -1 + \mu} 
    &= \int_0^\infty x_1^{-1+\mu}\ 
    \d x_1 \int_0^\infty \d x_2 \ x_2^{-1+\mu}\ 
    \psi^{(0)}_{2,\text{flat}}(x_1 + X_1, x_2 + X_2;Y)
    \nn\\
    &= \int_0^\infty \d x_1 \ x_1^{-1+\mu}\ \frac{\psi^{(0)}_{2,\textrm{flat}}(x_1+X_1,X_2;Y)}{\mu} + \mathcal{O}(\mu^{0})
    \nn\\
    &= \frac{\psi^{(0)}_{2,\textrm{flat}}(X_1,X_2;Y)}{\mu^2} + \mathcal{O}(\mu^{-1})~.
    \label{eq:flatspacelimitofFRWuplift}
\end{align}
Thus, for the flat-space case, the integral (\ref{eq:2-siteFRW}) returns the original wavefunction coefficient multiplied by a divergent factor $1/\mu^2$. 
This factor corresponds to the infinite volume of the energy integrals that we should not have integrated over.

The main objective of the paper is to illustrate how one can study the analytic structure of FRW wavefunction coefficients using techniques from \textit{relative twisted cohomology}, \textit{intersection theory} and canonical differential equations that has already been developed in context to flat-space scattering amplitudes. In particular, in the subsequent sections, we will derive the differential equations for the tree-level two-site/four-point and three-site/five-point and the one-loop two-site/two-point FRW wavefunction coefficients in a theory comprised of conformally coupled scalars obeying a $\lambda_3 \phi^3$ interaction (\ref{eq:actioninFRWphi^3theory}).
In the process, we argue that \textit{relative twisted cohomology} provides a particularly efficient and powerful formalism for studying FRW wavefunction coefficients. 


\section{The tree-level two-site FRW correlator \label{sec:4pt}}

\begin{figure}
    \centering
    \includegraphics[width=.3\textwidth]{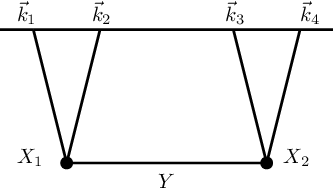}
    \caption{The Feynman diagram for the tree-level two-site/four-point wavefunction coefficient in $\lambda_3 \phi^3$ theory.}
    \label{fig:2sites}
\end{figure}

For the $\lambda_3 \phi^3$ theory of conformally coupled scalars (\ref{eq:actioninFRWphi^3theory}), the two-site flat-space wavefunction coefficient at tree-level is also the four-point function from the boundary perspective as depicted in figure \ref{fig:2sites}.
With the uplift prescription laid out in section \ref{sec:FRWcorr}, the corresponding FRW wavefunction coefficient has the integral representation
\begin{align} \label{eq:2siteFRW}
    \psi_{2,\text{FRW}}^{(0)}(X_1,X_2;Y) = \int_0^{\infty} \d x_1 \wedge \d x_2 \ \frac{(x_1 x_2)^{\vep}}{(x_1{+}x_2{+}X_1{+}X_2) (x_1{+}X_1{+}Y) (x_2{+}X_2{+}Y)}~.
\end{align}
While it is possible to integrate the above two-site/four-point integral, direct integration becomes intractable at higher points. 
Similar integrals often appear in the study of dimensionally regulated Feynman integrals where there is a rich literature describing their evaluation and understanding their analytic structure (see 
\cite{Eden:1966dnq,
Chetyrkin:1981qh,
Bern:1994cg,
Gehrmann:1999as,
Laporta:2000dsw,
Britto:2004nc,
Smirnov:2006ry,
Spradlin:2011wp,
Smirnov:2012gma,
Duhr:2014woa,
Gardi:2022wro,
Mizera:2021icv,
Mizera:2021fap,
Borinsky:2023jdv,
Matsubara-Heo:2023ylc} 
for a very incomplete list). 
In particular, we will use the method of canonical differential equations \cite{Henn:2013pwa, Henn:2014qga} and some ideas from intersection theory 
\cite{Mastrolia:2018uzb, Frellesvig:2019kgj, Mizera:2019gea, Mizera:2019vvs, Frellesvig:2020qot, Weinzierl:2020xyy, Chestnov:2022xsy, Caron-Huot:2021xqj, Caron-Huot:2021iev} 
to evaluate the two-site FRW correlator \eqref{eq:2siteFRW}.

In order to implement these methods, we have to consider the  family of integrals defined by the singularities of the integrand in \eqref{eq:2siteFRW}.
This family is made up of integrals whose integrands are rational differential forms on the manifold $X$ (to be defined later) multiplied by a universal multi-valued function, $u$, called the twist
\begin{align} \label{eq:4ptfam}
    \psi_{2,\text{FRW}}^{(0)}(X_1,X_2;Y;\bs{\mu},\bs{\nu}) &=
    \int u\ 
    \frac{
        N(\mbf{x})
    }{
        \prod_{j=1}^2 T_j^{\mu_j}
        \prod_{k=1}^3 S_k^{\nu_k}(\mbf{x})
    }
    \d^2\mbf{x}
    \quad
    (\nu_i,\mu_i \in \{0,1,2,\dots\}),
    \\ \label{eq:4ptTwist}
    u &= \prod_{i=1}^2 T_i^\vep
    \quad
    \text{
    ($\vep\notin\mathbb{C}\setminus\mathbb{Z}$ for FRW cosmologies)
    }
    .
\end{align}
In particular, the integral \eqref{eq:2siteFRW} corresponds to $\mu=\{0,0\}$ and $\nu=\{1,1,1\}$ above.  
While we can have any arbitrary polynomial numerator $N(\mbf{x})$, the analytic structure of the integral is (mostly) determined by the singular surfaces $T_a=0$ and $S_a=0$. 
Moreover, the behavior of the integrand at the singular surfaces $T_a=0$ and $S_a=0$ is very different and needs distinct mathematical treatment. 
The hyperplanes\footnote{By a hyperplane, we mean a codimension-1 surface cut out by a linear equation. We will use this terminology even for lines in 2-dimensions and planes in 3-dimensions.}
\begin{align}
    T_1 = x_1
    \quad\text{and}\quad
    T_2 = x_2~,
\end{align}
correspond to branch surfaces or \emph{twisted singularities} of the integrand. These singularities are said to regulated by $\vep$ and are mild in the sense that the integral in the neighbourhood of any $T_a=0$ has a well defined analytic continuation. 
On the other hand, the (hyper-)planes
\begin{align}
    S_1 & = x_1 + X_1 + Y~,
    \\
    S_2 & = x_2 + X_2 + Y~,
    \\
    S_3 & = x_1 + x_2 + X_1 + X_2~.
\end{align}
correspond to poles or \emph{relative singularities}. These singularities are dangerous and cannot be avoided by analytic continuation. 

Having understood the possible locations of singularities, we can define the integration manifold  $X=\mathbb{C}^2\setminus(T \cup S)$ where $T=\cup_{i=1}^2 \{T_i=0\}$ are the twisted singular surfaces and $S=\cup_{i=1}^3 \{S_i=0\}$ are the relative singular surfaces. The analytic structure of this family of integrals is almost entirely determined by how the surfaces $T_a$ and $S_a$ intersect amongst themselves. 

One perhaps surprising fact is that such families of integrals are spanned by a finite basis of integrals $\{I_a\}$ \cite{Lee:2013hzt, Bitoun:2018afx}. 
This means that the partial derivatives of these integrals with respect to the kinematic (non-integration) variables can be expressed as a linear combination of the original basis of integrals
\begin{align}
    \partial_z I_a = B^{(z)}_{ab} I_b~,
\end{align}
where $z\in\{X_1, X_2, Y\}$ for the example at hand. Such systems of differential equations have a closed solution in terms of a path-ordered exponential. 
Then, since we can can choose a basis such that the differential equation is proportional to $\vep$: $\mat{B} = \vep\mat{A}$, the path-ordered exponential 
\begin{align}
    \mbf{I} = \mathbf{I}_0 
    + \vep \int \mat{A} \cdot \mathbf{I}_0
    + \vep^2 \int \mat{A} \cdot \int \mat{A} \cdot \mathbf{I}_0 
    + \cdots 
    \label{eq:DEQsolpoe}
\end{align}
can be truncated at any desired order in $\vep$.\footnote{FRW integrals turn out to be sufficiently simple that it is always possible to chose such a basis.}
For each order in $\vep$, sophisticated technology exists in the amplitudes literature for handling the resulting iterated integrals \cite{bams/1183539443}. 
Lastly, $\mbf{I}_0$ corresponds to the boundary conditions of our basis integrals. 

Thus, we can integrate by differentiation!
Since obtaining $\mat{A}$ and the boundary conditions $\mbf{I}_0$ are orders of magnitude easier than direct integration, we pursue this strategy in the following sections. 

Before moving on, we note that the iterated integrals appearing in \eqref{eq:DEQsolpoe} will not produce anything more complicated than multiple polylogarithms because the matrix $\mat{A}$ contains at most simple poles (see \cite{Spradlin:2011wp, Duhr:2014woa, Duhr:2019tlz, Gardi:2022wro} for some of the mathematical techniques used to simplify such integrals). If one was powerful enough to sum the series \eqref{eq:DEQsolpoe}, one would recover the corresponding generalized hypergeometric function.
Further, note that $\vep$ carries weight $-1$ so that the expansion of these hypergeometric functions have uniform transcendental weight.

To aid us in the computation of differential equations, we introduce a few ideas from intersection theory. 
In section \ref{sec:4ptBad}, we introduce the twisted cohomology of FRW integrals, compute the differential equation and obtain the basis size from geometric quantities. Along the way we will comment on some computation bottlenecks of the usual twisted cohomology picture. To simplify the computation of the differential equation as well as improve our geometric understanding of such integrals, we introduce the vector space dual to FRW integrals in section \ref{sec:4ptSimp}. 
In particular, the analogous algorithm for computing the DEQ presented in section \ref{sec:4ptBad} simplifies and in some cases becomes trivial in the dual space.  
We also describe how to compute differential equations of logarithmic forms using the intersection number (an inner product on the FRW cohomology) bypassing integration-by-parts in section \ref{sec:4ptNoIBP}. 
In particular, we describe how to compute the differential equations without leaving the space of logarithmic forms.
Lastly, in section \ref{sec:4ptIntegration}, we will integrate the differential equation found in both sections \ref{sec:4ptBad} and \ref{sec:4ptSimp} to obtain an explicit formula for the two-site/four-point FRW correlator.

\subsection{Differential equations from twisted cohomology \label{sec:4ptBad}}

To compute the differential equations, it is useful to change perspective and consider the \emph{integrands} as the fundamental objects of interest instead of the corresponding integrals.\footnote{Since we will always be dealing with a basis of integrals there is no loss of information by considering integrands instead of integrals.}
The allowed FRW integrands are any holomorphic rational differential top-forms
$\vphi$ on $X$,\footnote{Holomorphic differential forms only contain the holomorphic variables of the complex manifold $X$. A top form has form degree equal to the complex dimension of the manifold $\text{dim}_\mathbb{C}X$. In this example, $\text{dim}_\mathbb{C}X = 2$.} 
multiplied by the multi-valued twist
\begin{align}
   \text{FRW integrands:}\ u\ \vphi.
\end{align} 
Mathematically, we denote the space of (holomorphic) differential $p$-forms by $\Omega^p(X)$. 
Since the twist is universal to all FRW integrands, we can factor it out and only work with the single-valued forms $\vphi$ by introducing a covariant derivative: $\d( u\ \vphi ) = u \nabla \vphi$ where $\nabla = \d + \omega \wedge$ and $\omega = \d\log u$ is a flat connection ($\d\omega=0$). The connection $\omega$ is analogous to the gauge field $A_\mu$ of QED and keeps track of the phase of $u$. 

Notice that integrands are not uniquely defined. One can always shift an integrand by a total derivative without changing the resulting integral
\begin{align}
    \int u\ \vphi 
    = \int \Big( u\ \vphi + \d (u\ \psi) \Big)
    = \int u \Big( \vphi +  \nabla \psi \Big)
\end{align}
for any $\psi\in\Omega^{\text{dim}_\mathbb{C}X-1}(X)$.\footnote{In the example at hand, $\text{dim}_\mathbb{C}X-1 = 2-1 = 1$.} 
Thus, the unique objects are holomorphic top-forms modulo covariant derivatives. 
This is is made precise by the \emph{twisted cohomology group} 
\begin{align}
    [\vphi] \in H^p(X,\nabla)
    = \frac{
        \{\vphi\in\Omega^n(X)\vert \nabla\vphi=0\}
    }{
        \{\nabla\psi \vert \psi\in\Omega^{p-1}(X)\}
    }.
\end{align}
The numerator corresponds to ``covariantly closed'' differential forms where the condition $\nabla\vphi=0$ ensures that the integral of $u\vphi$ does not depend on the choice of path -- only on the end points. The denominator corresponds to the set of all possible ``covariantly exact'' forms (covariant derivatives) that we can add to $\vphi$ without changing the resulting integral. 
We will use the freedom to add any $\nabla\psi$ to $\vphi$ to derive the FRW differential equations. In the Feynman integral literature, this procedure often goes by the name \emph{integration-by-parts}. 

\begin{figure}
    \centering
    \includegraphics[width=.4\textwidth]{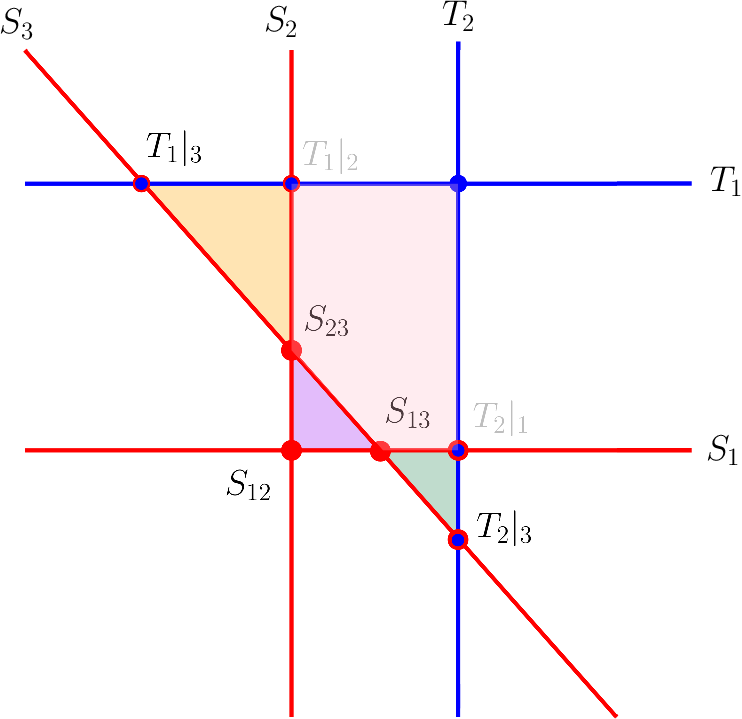}
    \caption{Hyperplane arrangement for the two-site/four-point integrand. The twisted planes are pictured in blue while the untwisted planes are red. For the FRW-cohomology, it is common to use the canonical forms of the bounded chambers, which are colored in orange, purple, green and pink (section \ref{sec:4ptBad}). For the dual-cohomology, the basis is determined by uplifting the basis on each 1- and 2-boundary since the bulk cohomology is empty. On each 2-boundary, the cohomology is spanned by the constant function. Each 1-boundary only has a non-trivial cohomology if it crosses both twisted lines (section \ref{sec:4ptSimp}). 
    }
    \label{fig:2siteHyperPlanes}
\end{figure}

Since twisted cohomology is well studied, we can borrow a key fact from the mathematical literature: only the top-dimensional twisted cohomology group is non-trivial.\footnote{The usual proof of these theorems assume some conditions that we break by letting the $S_i$ have integer exponents. Fortunately, in most cases, this statement remains true even when the $S_i$ have integer exponents. A proof of this fact will appear in a forthcoming publication by one of the authors.} 
As a consequence, the dimension of the top-dimensional cohomology (or equivalently the size of the FRW basis) is equal to the Euler characteristic $\chi(X)$. Moreover, it is known that for hyperplane arrangements the Euler characteristic is equal to the number of bounded chambers \cite{yoshida1997hypergeometric, aomoto2011theory}.\footnote{When all singularities are twisted, a more practical way to compute this Euler characteristic is by counting the critical points of the connection \cite{Lee:2013hzt, Mizera:2017rqa}. While this formula does not hold for FRW cohomology, it does hold for the dual cohomology. Thus, it is the recommend way for computing the dual basis size when working in higher dimensions.\label{foot:critPoints}} 
For our particular example, there are four bounded chambers (figure \ref{fig:2siteHyperPlanes}). This should be familiar from the positive geometry and cosmological polytope picture 
\cite{ArkaniHamed2017, Arkani-Hamed:2018bjr, Benincasa:2019vqr, Kuhne:2022wze, Juhnke-Kubitzke:2023nrj, Albayrak:2023hie}. 
In fact, the canonical forms associated to the bounded chambers provide a natural $\vep$-form basis for this family of integrals\footnote{The canonical form of a chamber bounded by the lines $A,B,C$ is given by the $\d\log$-form $\d\log \frac{A}{B} \wedge \d\log\frac{B}{C}$ or any permutation of $A,B$ and $C$. The canonical form for a chamber bounded by more than three lines is obtained by triangulating the bounded chamber and adding up the canonical forms associated to the triangulation. This construction generalizes straightforwardly to higher dimensional spaces.}
\begin{align} \label{eq:4ptCan}
    \color{DarkOrchid}
    \vth_1 &= 
        \d\log\frac{S_1}{S_2}
        \wedge\d\log\frac{S_2}{S_3},
    \\
    \color{Orange}
    \vth_2 &= 
        \d\log\frac{T_1}{S_2}
        \wedge\d\log\frac{S_2}{S_3},
    \\
    \color{Green}
    \vth_3 &= 
        \d\log\frac{T_2}{S_1}
        \wedge\d\log\frac{S_1}{S_3},
    \\
    \color{Lavender}
    \vth_4 &= \bigg(
        \frac{x_1+X_1+X_2}{T_2 S_1 S_3}
        + \frac{x_2+X_1+X_2}{T_1 S_2 S_3}
        - \frac{1}{T_1 T_2}
    \bigg) \d x_1 \wedge \d x_2.
\end{align}
The last form has been written out explicitly, since its construction requires one to triangulate the pink bounded chamber in figure \ref{fig:2siteHyperPlanes} and add up the canonical forms on each piece of the triangulation. 
Notice that $u\ \vth_1$ is simply the integrand of \eqref{eq:2siteFRW} (up to a sign).

Let us now compute the differential equation for $\vth_1$. Taking the kinematic covariant derivative of $\vth_1$ yields 
\begin{align} \label{eq:thetaDot}
    \dot{\vth}_1 \equiv \nabla_\text{kin} \vth_1
    &= \frac{
        2 \left(2 x_1+x_2+2 X_1+X_2+Y\right) 
    }{S_1^2 S_2 S_3^2} 
    \ \d x_1 \wedge \d x_2 \wedge \d X_1
    \nn\\&\quad
    + \frac{
        2 \left(x_1+2 x_2+X_1+2 X_2+Y\right) 
    }{S_1 S_2^2 S_3^2}
    \ \d x_1 \wedge \d x_2 \wedge \d X_2
    \nn\\&\quad
    + \frac{
        2 \left(x_2 X_1+x_1 X_2+x_1 x_2+X_1 X_2-Y^2\right) 
    }{S_1^2 S_2^2 S_3}
    \ \d x_1 \wedge \d x_2 \wedge \d Y,
\end{align}
where $\nabla_\text{kin} = \d_{\text{kin}} + \d_\text{kin} \log u = \d_{\text{kin}}$ and $\d_\text{kin} = \partial_{X_1} \d X_1 + \partial_{X_2} \d X_2 + \partial_Y \d Y$ for the current example. 
Note that $\dot{\vth}_1$ is not directly expressible in terms of our basis \eqref{eq:4ptCan}. However, we have the freedom to add the covariant derivative of any 1-form to \eqref{eq:thetaDot} without changing the value of the integral. One can check that 
\begin{align} \label{eq:Avth1}
    \dot{\vth}_1 + \nabla \text{IBP} 
    &= \vep\ \d\log \left(\left(X_1+Y\right) \left(X_2+Y\right)\right) \wedge \vth_1 
    \nn\\&
    + \vep\ \d\log \left(\frac{Y-X_1}{X_1+Y}\right)  \wedge \vth_2
    + \vep\ \d\log \left(\frac{X_2+Y}{Y-X_2}\right) \wedge \vth_3,
\end{align}
where 
\begin{align}
    \text{IBP} &= \frac{
        2 (x_1+X_1) \d x_2 \wedge \d Y
        -2 (x_2+X_2) \d x_1 \wedge \d Y
        +2 Y (
            \d x_1 \wedge \d X_2 
            - \d x_2 \wedge \d X_1
        )
    }{S_1 S_2 S_3}
    . 
\end{align}
The IBP form above is not so difficult to guess or construct for the current example. However, constructing such IBP forms at the level of the three-site/five-point graph or higher is much more difficult. This integration-by-parts step can be simplified and in some cases trivialized by working in the dual cohomology of section \ref{sec:4ptSimp}. While the change to the dual cohomology requires a new layer of abstraction, the computational and conceptual simplifications are well worth it! 

Equation \eqref{eq:Avth1} corresponds to the first row of our differential equation.
Computing the other rows of $\mat{A}$ in an analogous fashion yields 
\begin{align}
    \mat{A}_\vth \label{eq:Avth}
    = \vep\ \d
    {\tiny
    \begin{pmatrix}
        \log \left(\left(X_1+Y\right) \left(X_2+Y\right)\right) 
        & \log \left(\frac{Y-X_1}{X_1+Y}\right) 
        & \log \left(\frac{X_2+Y}{Y-X_2}\right) 
        & 0 
    \\
        0 
        & \log \left(\left(X_1-Y\right) \left(X_1+X_2\right)\right) 
        & \log \left(\frac{X_2+Y}{X_1+X_2}\right) 
        & \log \left(\frac{X_2+Y}{X_1+X_2}\right) 
    \\
        0 & \log \left(\frac{X_1+Y}{X_1+X_2}\right) 
        & \log \left(\left(X_2-Y\right) \left(X_1+X_2\right)\right) 
        & \log \left(\frac{X_1+X_2}{X_1+Y}\right) 
    \\
        0 & \log \left(\frac{Y-X_1}{X_1+Y}\right) 
        & \log \left(\frac{X_2+Y}{Y-X_2}\right) 
        & \log \left(\left(X_1+Y\right) \left(X_2+Y\right)\right)
    \end{pmatrix}
    },
\end{align}
where we have included a subscript $\vth$ on $\mat{A}$ to signify the choice of basis. Indeed, such differential equations come with an inherent gauge redundancy that is the choice of basis. Under a change of basis $\vth_i = U_{ij} \vphi_j $, the matrix $\mat{A}$ transforms like a gauge field 
\begin{align}  
    \nabla_\text{kin} \bs{\vphi}
    = \mat{A}_\vphi \cdot  \bs{\vphi},
\end{align}
where 
\begin{align}
    \mat{A}_\vphi = \mat{U}^{-1} \cdot \mat{A}_\vth \cdot \mat{U} 
    - \mat{U} \cdot \d_\text{kin} \mat{U}.
\end{align}
The equation above gives the rule for how to compare the differential equations for different choices of bases. 
Notice that \eqref{eq:Avth} is actually quite dense. This hints that there may be a better choice of basis that simplifies the form of $\mat{A}$. In fact, the dual cohomology will point us to such a  basis in section \ref{sec:4ptSimp}. 

Before moving onto describe the dual cohomology, we note that the matrix $\mat{A}$ is integrable: $\d_\text{kin} A_{ij} + A_{ik} \wedge A_{kj} = 0$. This condition places strong consistency conditions on $\mat{A}$ and provides an important check of our calculations. 
If $\mat{A}$ were not integrable, the path ordered exponential would depend on the path of integration in kinematic space.

\subsection{Differential equations from the dual (relative twisted) cohomology \label{sec:4ptSimp}}

Since the twisted cohomology is a vector space, there exists a dual vector space and an associated inner product called the \emph{intersection number}. The dual cohomology is the \emph{relative twisted} cohomology $\c{H}^p(X,\nabla) = H^p(\c{X}, \c{X} \cap S,\c{\nabla})$ where $\c{X} = \mathbb{C}^2 \setminus T$, $\c{\nabla}=\d+\com\wedge$ is the dual covariant derivative, $\com=\d\log\cu$ is the dual connection and $\cu=1/u$ is the dual twist \cite{Matsumoto:aa, Caron-Huot:2021xqj, Caron-Huot:2021iev}. In this section, we will unpack the ideas underlying relative twisted cohomology and describe how this simplifies computing differential equations. 


We also opt to avoid delving into the details of how the intersection number is computed. Instead, we give a simple formula for the intersection of logarithmic forms in appendix \ref{app:intNum} that covers all cases considered in this work. However, to provide some motivation for why the dual cohomology takes its particular form, it is useful to give the definition of the intersection number in the main text. 

The intersection number of a FRW-form $\vphi \in H^n(X,\nabla)$ with a dual form $\cvphi\in H^n(\c{X},\c{X} \cap S,\c{\nabla})$ is given by
\begin{align} \label{eq:intNumDef}
    \la \cvphi \vert \vphi \ra
    \propto \frac{1}{ (2 \pi i)^{\text{dim}_\mathbb{C} X} } \int_{X} \text{Reg}[\cvphi] \wedge \vphi
\end{align}
where the overall sign is conventional and $\text{Reg}[\cvphi]$ is a regulated version of the dual form.\footnote{We skip the details of the regularization procedure in this work since the simple formula in appendix \ref{app:intNum} covers all cases encountered in this work. For details on the regularization procedure we direct the interested reader to \cite{matsumoto1998kforms, Mizera:2017rqa, Mizera:2019gea, Mizera:2019vvs, Frellesvig:2019uqt, Weinzierl:2020xyy, Chestnov:2022xsy, Caron-Huot:2021iev, Caron-Huot:2021xqj} and the references therein.} 
While this looks like it could be a difficult integral to perform, the intersection number  localizes on the maximal codimension intersections of the singular surfaces $T_i$ and $S_i$ (the point $S_{12}$ in figure \ref{fig:2sites} for example).
Thus, the intersection number is actually algebraically evaluated as a series of residues. 
Moreover, it is simple to compute for logarithmic forms, which encompass all FRW-forms. 

Like any vector space, we can construct a resolution of the identity from a basis for the FRW and dual cohomologies $\{\vphi_a\}$ and $\{\cvphi_a\}$
\begin{align}
    \mathds{1} = 
    \sum_{a,b} \vert\vphi_a\ra \ C_{a b}^{-1} \ \la \cvphi_b \vert
    .
\end{align}
Here, $C_{ab} = \la \cvphi_a \vert \vphi_b \ra$ is the intersection matrix associated to our choice of bases.
Using this resolution of identity, any FRW-form $\vphi$ (and hence integral $\int u\ \vphi$) can be projected onto the basis $\{\vphi_a\}$
\begin{align}
    \vert \vphi \ra = \sum_a c_a \vphi_a,
    \qquad
    c_a = C_{ab}^{-1} \la \cvphi_b \vert \vphi \ra.
\end{align}
Roughly speaking, this projection is a form of generalized unitary \cite{Bern:1994cg, Bern:2004ky, Britto:2004nc, Anastasiou:2006jv, Britto:2006fc, Ossola:2006us, Britto:2007tt, Bourjaily:2017wjl, Feng:2021spv}.
It can also be used instead of the integration-by-parts identities of the previous section to derive the differential equations (section \ref{sec:4ptIntegration}).

The reason that the dual cohomology takes a different form from the FRW cohomology is due to the nature of the regularization map in \eqref{eq:intNumDef}. Since the twisted singularities $T_a$ can be regularized thanks to the fact that $\vep \notin \mathbb{Z}$ is generic, dual forms can have singularities at the zero loci of the $T_i$. On the other hand, singularities at the zero loci of the relative singularities $S_i$ \emph{cannot} be regularized. Thus, dual forms must be regular on $\c{X} \cap S$. This is why we only excise $T$ in the definition of the dual manifold $\c{X} = \mathbb{C}^2 \setminus T$.

The last piece of mathematical terminology that we need to explain before moving to the calculation of the differential equation is what the adjective ``relative'' means. It is simply a fancy name for keeping track of boundary terms that arise form Stokes' theorem. For the FRW cohomology, we did not allow boundaries and the total covariant derivative always integrated to zero. For the dual cohomology, the relative singular surfaces become boundaries and the integral of a total covariant derivative only vanishes up to boundary terms. These boundary terms are integrals arising from lower-degree forms that live on the boundary surfaces. 

Mechanically, this boundary information is kept track by the so-called coboundary symbol $\delta_J$. A generic dual form $\cvphi \in H^n(\c{X},\c{X} \cap S,\c{\nabla})$ should be thought of as a vector with components labeled by each boundary $J$
\begin{align}
    \cvphi = \sum_J \delta_J\left( \cphi_{J} \right).
\end{align} 
Here, $\delta_J$ can be thought of as picking out the component of $\cvphi$ that corresponds to the boundary $S_J=\c{X}\cap_{j\in J}S_j$ and $\cphi_{J}$ is an element of the twisted cohomology of the boundary indexed by $J$: $H^{n-|J|}(\c{X} \cap S_J, \c{\nabla})$.\footnote{For the mathematically inclined, the formal definition of relative twisted cohomology is the direct sum of the twisted cohomology on each boundary 
\begin{align}
    \Omega^n(\c{X},S; \c{\nabla}) 
    &=
    \Omega^n(\c{X}; \c{\nabla})
    \ \bigoplus_{a=1}^m \ 
    \Omega^{n-1}(\c{X} \cap S_a; \c{\nabla}\vert_a)
    \ \underset{a\neq b}{\bigoplus_{a,b=1}^m} \ 
    \Omega^{n-2}(\c{X} \cap S_a \cap S_b; \c{\nabla}\vert_{a b})
    \nn\\
    &\quad
    \bigoplus\quad \cdots 
    \underset{a_i \neq a_j}{\bigoplus_{a_1,\dots,a_n=1}^m} \Omega^{0}(\c{X} \cap S_{a_1} \cap \cdots \cap S_{a_n}; \c{\nabla}\vert_{a_1 \cdots a_n})
    \nn
\end{align}
where $m$ is the number of relative surfaces $S_i$.
} 
Naturally, the corresponding dual integrals
\begin{align}
    \c{I}_J 
    = \int_\gamma \cu\ \cvphi
    = \sum_J \int_{\gamma} \cu\ \delta_{J}\Big( \cphi_{J} \Big)
    = \sum_{J} \int_{\gamma\vert_J} \cu\vert_J\  \cphi_{J},
\end{align}
are simply integrals defined on the boundaries.

In addition to indexing the components of $\cvphi$, the $\delta_J$ act like differential forms. They anti-commute with themselves
\begin{align}
    \delta_{ij} &= \delta_i \wedge \delta_j = - \delta_j \wedge \delta_i = -\delta_{ji}
\end{align}
and the exterior derivative is given by 
\begin{align} \label{eq:bdCalc}
    \d \delta_{J}(\cphi) &= (-1)^{|J|} \delta_J\left(
        \d \cphi
    \right)
    + (-1)^{|J|} \sum_{k\notin J} \delta_{Jk}\left(
        \cphi\vert_k
    \right)
\end{align}
up to boundary terms. The $\delta_J$ also convert the covariant derivative on bulk space to the covariant derivative on the boundary 
\begin{align} \label{eq:parallelTransport}
    \c{\nabla} \delta_J( \cphi )
    = (-1)^{|J|} \delta_J\left( \c{\nabla}\vert_J\ \cphi \right)
    + (-1)^{|J|} \sum_{k\notin J} \delta_{Jk}\left(  \cphi\vert_k \right)
\end{align}
where $\c{\nabla}\vert_J = \d + \com\vert_J \wedge$. Equations \eqref{eq:bdCalc} and \eqref{eq:parallelTransport} will do virtually all the heavy lifting in this formalism. To the zeroth order approximation, the $\delta_i$ can be thought of as being dual to the $\d\log$-form $\d\log S_i$ since the $\delta_i$ correspond to residues about $S_i=0$ in the intersection number (see equation \eqref{eq:deltaInt}). 
The main job of the $\delta_J$ is to take care of the combinatorics of Stokes' theorem on a manifold with boundary.

\begin{figure}
    \centering
    \includegraphics[width=.5\textwidth]{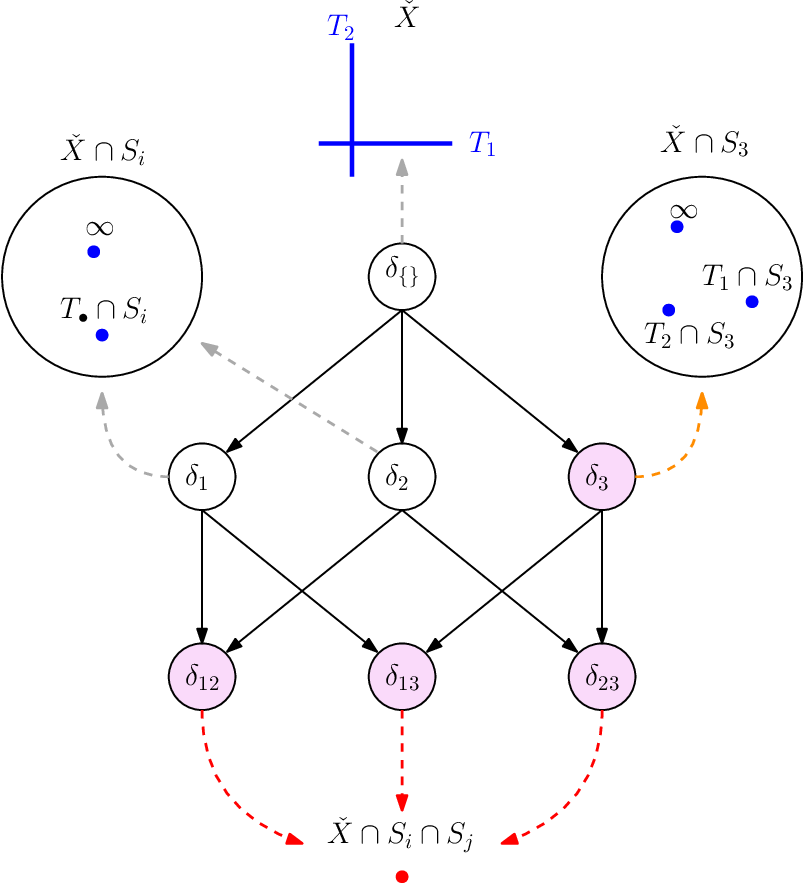}
    \caption{The topological spaces appearing in the direct sum decomposition of the relative cohomology. 
    Since the twist is sufficiently generic, only the corresponding middle dimensional cohomology is non-trivial. 
    The first diagram is a real slice of the bulk space where the coordinate planes (in blue) have been removed. Since the planes defined by the $T_{i}$ do not form a bounded chamber, $\dim H^{2}(\c{X},\c{\nabla}) = 0$.
    Each 1-boundary $\c{X} \cap S_i$ is topologically a Riemann sphere with either 2- or 3-punctures. Only $X \cap S_3$ has enough punctures to have a non-trivial topology. Thus, $\dim H^1(\c{X} \cap S_3,\c{\nabla}) = 1$ while $\dim H^1(\c{X} \cap S_{i=1,2},\c{\nabla}) = 0$.
    The last diagram represents the 2-boundaries $\c{X} \cap S_i \cap S_j$. While there is no twisting, the $0^\text{th}$ cohomology of a point is known to be 1-dimensional: $\dim H^0(\c{X} \cap S_1 \cap S_j) = 1$.
    }
    \label{fig:4pt_top}
\end{figure}

Returning to the two-site/four-point example, we need to choose a basis for the dual-cohomology. To do this, we simply write down a basis for the cohomology on each boundary. First, we look at the bulk twisted cohomology $H^2(\c{X},\c{\nabla})$. The topological space $\c{X}$ is pictured at the top of figure \ref{fig:4pt_top}. Clearly there are no bounded chambers cut out by the twisted singularities so  $\text{dim}H^2(\c{X},\c{\nabla}) = 0$. Next, we look at the cohomologies on all possible 1-boundaries. We find two distinct cases of 1-boundaries (pictured on the left and right in figure \ref{fig:4pt_top}). Both boundaries $S_{1}$ and $S_{2}$ are twice punctured Riemann spheres. From a well known result \cite{aomoto2011theory},  the dimension of the twisted cohomology of a punctured Riemann sphere is $\#(\text{punctures})-2$ (this is the analogue of counting bounded chambers in 1-dimension). Thus, both $H^1(S_1,\c{\nabla}\vert_1)$ and $H^1(S_2,\c{\nabla}\vert_2)$ are trivial. On the other hand, the twisted cohomology on $S_3$ is 1-dimensional since $S_3$ has 3 punctures. We can choose 
\begin{align} \label{eq:4pt_1bd_basis}
	\cphi_1 = \vep\ \d\log \frac{T_1\vert_3}{T_2\vert_3}
\end{align}
for the basis of  $H^1(S_3,\c{\nabla}\vert_3)$ and then uplift this into the full dual cohomology by simply tacking on a $\delta_3$
\begin{align}
	\cvphi_1 
	= \delta_3\left(\cphi_1\right) 
	= \vep\ \delta_3\left(\d\log \frac{T_1\vert_3}{T_2\vert_3}\right). 
\end{align}
The 1-boundary form has been normalized by $\vep$ to ensure that the differential equation is in $\vep$-form.\footnote{Intuitively, the more a form is localized to a boundary the less transcendental the resulting integral is. Thus, we normalize the more transcendental 1-boundary form by $\vep$ so that it has the same transcendentality as the 2-boundary forms. }

Lastly, there are three 2-boundaries: $S_{12},S_{23}$ and $S_{13}$. Each of these are simply a point (bottom of figure \ref{fig:4pt_top}). While the corresponding cohomology $H^0(S_{ij})$ is not twisted, it is non-trivial and spanned by the constant function: 1. Thus, we choose 
\begin{align} \label{eq:4pt_2bd_basis}
    \cvphi_2 = \delta_{12}(1),
    \qquad
    \cvphi_3 = \delta_{23}(1),
    \qquad
    \cvphi_4 = \delta_{13}(1),
\end{align}
as the remaining elements for our basis of the dual cohomology. Notice that the basis size is the same as in section \ref{sec:4ptBad}. In the dual cohomology, the basis size is computed by adding the dimensions of the twisted cohomology on each possible boundary!\footnote{Sometimes this can over count the basis size when the hyperplane arrangement is non-generic and some boundaries need to be identified. We will see an example of this in section \ref{sec:5pt}.}
For higher dimensional manifolds, one can no longer visualize and count bounded chambers and we have to switch to alternative methods for computing the Euler characteristic of each boundary. 
A practical way to do this is provided by footnote \ref{foot:critPoints}.

The components of the kinematic connection associated to the 2-boundary forms are trivial to obtain. Taking the (dual) kinematic covariant derivative and using equation \eqref{eq:parallelTransport}, we find
\begin{align}
    \c{\nabla}_\mathrm{kin} \cvphi_2
    &= \delta_{12}\left(\com_\mathrm{kin}\vert_{12}\right)
    = \com_\mathrm{kin}\vert_{12} \wedge \cvphi_2
    \equiv \c{A}_{22} \wedge \cvphi_2,
    \\
    \c{\nabla}_\mathrm{kin} \cvphi_3 
    &= \delta_{23}\left(\com_\mathrm{kin}\vert_{23}\right)
    = \com_\mathrm{kin}\vert_{23} \wedge \cvphi_3
    \equiv \c{A}_{33} \wedge \cvphi_3,
    \\
    \c{\nabla}_\mathrm{kin} \cvphi_4
    &= \delta_{13}\left(\com_\mathrm{kin}\vert_{13}\right)
    = \com_\mathrm{kin}\vert_{13} \wedge \cvphi_4
    \equiv \c{A}_{44} \wedge \cvphi_4,
\end{align}
where $\c{\nabla}_\text{kin} = \d_\text{kin} + \com_\text{kin}$, $\com_\text{kin} = \d_\text{kin}\log\cu$ and $\com_\text{kin}\vert_J = \d_\text{kin}\log\cu\vert_J$.
Consequently, we see that the dual differential equation has the following upper triangular form
\begin{align}
    \c{\nabla}_\mathrm{kin} \cvphi_a = \c{A}_{ab} \wedge \cvphi_b
    \quad \text{where} \quad
    \mat{\c{A}} = \begin{pmatrix}
        \c{A}_{11} & \c{A}_{12} & \c{A}_{13} & \c{A}_{14}
        \\
        0 & \com_\mathrm{kin}\vert_{12} & 0 & 0
        \\
        0 & 0 & \com_\mathrm{kin}\vert_{23} & 0
        \\
        0 & 0 & 0 & \com_\mathrm{kin}\vert_{13}
    \end{pmatrix}
    .
\end{align}

To compute the remaining components associated to the 1-boundary form, we perform IBP on the 1-boundary. That is, we look for some kinematic 1-forms $A_{1i}$ such that 
\begin{align} \label{eq:4pt A1i}
    \c{\nabla}_\mathrm{kin} \cvphi_1
    + \c{\nabla} \ibp
    = \sum_{i=1}^4 \c{A}_{1i} \wedge \cvphi_i,
\end{align}
for some integration-by-parts form $\ibp$. 
Since $\cvphi_1$ is a form on the boundary $S_3$, $\ibp$ should also be a form on this boundary: $\ibp = \delta_{3}\left(\c{\xi}\right)$ where $\c{\xi}$ is a 1-form in the kinematic variables but a 0-form in the integration variables. 
Inserting this definition of $\ibp$ into \eqref{eq:4pt A1i}, we obtain
\begin{align} \label{eq:IBPConstraint}
    &\delta_3\left( 
        \c{A}_{11} \wedge \cphi_1
        - \c{\nabla}_{\mathrm{kin}}\vert_3 \ \cphi_1
        - \c{\nabla}_{\mathrm{int}}\vert_3 \ \c{\xi}
    \right)
    \nn\\&\qquad
   + \delta_{12}\left(
        \c{A}_{12} 
   \right)
   + \delta_{23}\left(
        \c{A}_{13} + \c{\xi}\vert_{2}
   \right)
   + \delta_{13}\left(
        \c{A}_{14}
        + \c{\xi}\vert_{1}
   \right)
   = 0~.
\end{align}
Again, we see that the 2-boundary components are fixed rather trivially
\begin{align}
    \c{A}_{12} = 0~,
    \quad
    \c{A}_{13} = - \c{\xi}\vert_{2}~,
    \quad
    \c{A}_{14} &= - \c{\xi}\vert_{1}~.
 \end{align}
All that remains is to specify $\c{\xi}$. 
It is easily verified that equation \eqref{eq:IBPConstraint} is satisfied by setting
$\c{A}_{11} = -2\vep\ \d\log(X_1+X_2)$ and 
\begin{align}
    \c{\xi} = -\vep \frac{\d (X_1 + X_2)}{T_2\vert_3}.
\end{align}
Putting everything together, we find the dual differential equation 
\begin{align} \label{eq:4ptDualDEQ}
    \mat{\c{A}} = -\vep\ \d
    {\tiny
    \begin{pmatrix}
        2\log(X_1+X_2) 
        &
        0
        &
        \log\left(\frac{X_2+Y}{X_1-Y}\right)
        & 
        \log\left(\frac{X_2-Y}{X_1+Y}\right)
        \\
        0
        & 
        \log(X_1+Y)(X_2+Y)
        & 
        0
        & 
        0
        \\
        0
        &
        0
        & 
        \log\left((X_1-Y)(X_2+Y)\right)
        &
        0
        \\
        0
        &
        0
        &
        0
        & 
        \log\left((X_2-Y)(X_1+Y)\right)
    \end{pmatrix}
    }
\end{align}
for our basis of the dual-cohomology \eqref{eq:4pt_1bd_basis} and \eqref{eq:4pt_2bd_basis}.

The differential equation for the basis of canonical forms \eqref{eq:Avth} is obtained from the dual differential equation \eqref{eq:4ptDualDEQ} by a gauge transform 
\begin{align}
    &\mat{A}_\vth = - \left(
        \mat{C}^{-1}_\vth \cdot \mat{\c{A}} \cdot \mat{C}_\vth
    \right)^\top
    \nn\\
    &= \vep\ \d
    {\tiny
    \begin{pmatrix}
        \log \left(\left(X_1+Y\right) \left(X_2+Y\right)\right) 
        & \log \left(\frac{Y-X_1}{X_1+Y}\right) 
        & \log \left(\frac{X_2+Y}{Y-X_2}\right) 
        & 0 
    \\
        0 
        & \log \left(\left(X_1-Y\right) \left(X_1+X_2\right)\right) 
        & \log \left(\frac{X_2+Y}{X_1+X_2}\right) 
        & \log \left(\frac{X_2+Y}{X_1+X_2}\right) 
    \\
        0 & \log \left(\frac{X_1+Y}{X_1+X_2}\right) 
        & \log \left(\left(X_2-Y\right) \left(X_1+X_2\right)\right) 
        & \log \left(\frac{X_1+X_2}{X_1+Y}\right) 
    \\
        0 & \log \left(\frac{Y-X_1}{X_1+Y}\right) 
        & \log \left(\frac{X_2+Y}{Y-X_2}\right) 
        & \log \left(\left(X_1+Y\right) \left(X_2+Y\right)\right)
    \end{pmatrix}
    }
\end{align}
where
\begin{align} \label{eq:canCmat}
    (C_\vth)_{ab} = 
    \la\cvphi_a\vert\vth_b\rangle
    = \begin{pmatrix}
        0 & -1 & 1 & 0 
        \\
        1 & 0 & 0 & 0
        \\
        1 & 1 & 0 & 1
        \\
        -1 & 0 & 1 & -1
    \end{pmatrix}
\end{align}
is the intersection matrix. 
The intersection matrix is analogous to the computation of generalized unitary coefficients in the context of Feynman integrals. In particular, the simple formula for the intersection number of $\d\log$-forms (appendix \ref{app:intNum}) make such calculations straightforward!   


Furthermore, the intersection matrix is a realization of the isomorphism between the FRW- and dual-cohomologies since it has full rank.
However, in some sense, the canonical forms \eqref{eq:4ptCan} are not the most direct way of establishing the duality between the FRW- and dual-cohomologies since the intersection matrix is dense. 

Instead, one can directly construct a basis for the FRW-cohomology that is orthonormal to the basis of dual-cohomology \eqref{eq:4pt_1bd_basis} and \eqref{eq:4pt_2bd_basis}.
By choosing $\d\log$ forms that do not share singularities on the maximal intersections of the hyperplanes $\{T_i,S_i\}$, one guarantees that the corresponding intersection matrix is diagonal. For example, the following basis 
\begin{align}
    \vphi_1 &= -
        \frac{1}{2} \d\log(S_3) 
        \wedge \d\log\left(\frac{T_1}{T_2}\right)~,
    &
    \vphi_2 &= \d\log S_1 \wedge \d\log S_2~,
    \nn\\
    \vphi_3 &= \d\log S_2 \wedge \d\log S_3~,
    &
    \vphi_4 &= \d\log S_1 \wedge \d\log S_3~,
\end{align}
is orthonormal to the $\cvphi_a$ 
\begin{align}
    (C_\vphi)_{ab} = \la \cvphi_a \vert \vphi_b \ra = \delta_{ab}~.
\end{align} 

Generally, it is much easier to build a basis in the dual-cohomology that has a simple DEQ. Thus, the basis $\vphi_a$ is extremely useful since the resulting differential equation is the minus transpose of $\mat{\c{A}}$
\begin{align} \label{eq:Avphi}
    \mat{A}_\vphi = - \left(\mat{\c{A}}\right)^\top,
\end{align}
which is much more sparse than $\mat{A}_\vth$. 
Then, since the integral of physical interest is simply related to the $\vphi$-basis ${\color{DarkOrchid}\vth_1} = \vphi_2 + \vphi_3 - \vphi_4$\footnote{These signs simply correspond to the first column of $\mat{C}_\vth$. By inserting identity $\mathds{1} = \vert\vphi_a\ra \left(C_\vphi^{-1}\right)_{ab} \la\cvphi_b\vert$, one finds $\vert\vth_1\ra = \vert\vphi_a\ra \left(C_\vphi^{-1}\right)_{ab} \la\cvphi_b\vert\vth_1\ra = \la\cvphi_a\vert\vth_1\ra  \vert\vphi_a\ra = \left(C_\vth\right)_{a1} \vert\vphi_a\ra$ as claimed.}, one can integrate the simpler DEQ $\mat{A}_\vphi$ instead of $\mat{A}_\vth$. 
Moreover, the $\vphi$-basis never needs to be explicitly constructed -- only its existence is important! 
One can obtain any FRW integral of interest by integrating $\left(-\mat{\c{A}}\right)^\top$ (without choosing a basis for the FRW cohomology) and using the intersection matrix to project out the correct linear combination.

\subsection{Trivializing the DEQ computation using intersection theory \label{sec:4ptNoIBP}}

As advertised in the introduction, one can compute the differential equations for a family of relative twisted differential forms using the intersection number instead of IBPs. Explicitly, 
\begin{align} \label{eq:DEQviaInt}
    A_{ij} = C^{-1}_{jk} \la \cvphi_k \vert \nabla_{\text{kin}} \vphi_i \ra \,, 
    \qquad
    \check{A}_{ij} = \la \check{\nabla}_{\text{kin}} \cvphi_i \vert \vphi_k \ra C^{-1}_{kj} 
    \,,
\end{align}
where $C_{ij} = \la \cvphi_i \vert \vphi_j \ra$ is the intersection matrix for a given choice of basis.
In general, this usually requires more sophisticated methods for computing the intersection numbers than those presented in appendix \ref{app:intNum} since the forms $\nabla_{\text{kin}} \vphi_j$ and $\check{\nabla}_{\text{kin}} \cvphi_i$ no longer have at most simple poles. However, since our forms are logarithmic, we can use a simple trick that avoids the introduction of double poles and allows us to compute the differential equations using only the formula presented in appendix \ref{app:intNum}!\footnote{The methods outlined here also help streamline the integration-by-parts algorithm presented in sections \ref{sec:4ptBad} and \ref{sec:4ptSimp}.}

The idea is to simply promote the exterior derivative $\d$ that acts on the integration variables to the exterior derivative that acts on both the integration and external variables: $\d \to \D = \d + \d_\text{kin}$. This promotes our basis to forms valued on the total space. For example, 
\begin{align}
    {\color{DarkOrchid} \vth_1} 
    &\to \D\log\frac{S_1}{S_2} 
    \wedge  \D\log\frac{S_2}{S_3}
    \nn\\&
    =\frac{- 2 Y\ \d x_1\wedge \d x_2}{(x_1+x_2+X_1+X_2) (x_1+X_1+Y) (x_2+X_2+Y)}
    \nn\\&\quad
    +\frac{2(x_2+X_2)\ \d x_1\wedge \d Y}{(x_1+x_2+X_1+X_2) (x_1+X_1+Y) (x_2+X_2+Y)}
    \nn\\&\quad
    +\frac{2Y\ \d x_2\wedge \d X_1}{(x_1+x_2+X_1+X_2) (x_1+X_1+Y) (x_2+X_2+Y)}
    \nn\\&\quad
     -\frac{2Y\ \d x_1\wedge \d X_2}{(x_1+x_2+X_1+X_2) (x_1+X_1+Y) (x_2+X_2+Y)}
     \nn\\&\quad
    -\frac{2(x_1+X_1)\ \d x_2\wedge \d Y}{(x_1+x_2+X_1+X_2) (x_1+X_1+Y) (x_2+X_2+Y)}
    \nn\\&\quad
    + \text{kinematic 2-forms}
    \,.
\end{align}
Note that the FRW connection does not change since it is independent of kinematics $\omega \to \D\log(u) = \d\log(u)$.\footnote{On the other hand, the dual connection does change since the restriction to boundaries introduces kinematic dependence.}

Now, since the difference between $(\D+\D\log(u)\wedge)$ and $\nabla_\text{kin}$ is a total derivative $\nabla$, we can replace $\nabla_{\text{kin}}$ in \eqref{eq:DEQviaInt} by $(\D+\D\log(u)\wedge)$ without changing the intersection number.\footnote{The intersection number of any total covariant derivative $\nabla$ or $\c{\nabla}$ always vanishes.} 
Explicitly, for $\color{DarkOrchid}\vth_1$, this amounts to the replacement 
\begin{align}
    {\color{DarkOrchid}\dot{\vth_1}} 
    = \nabla_\text{kin} {\color{DarkOrchid}\vth_1} 
    &\to (\D+\D\log(u)\wedge) 
    \ {\color{DarkOrchid}\vth_1}
    \nn\\&
    = \frac{- 2\vep\  (x_2 X_1+x_1 X_2+2 x_1 x_2)\ \d x_1\wedge \d x_2\wedge \d Y}{x_1 x_2 (x_1+x_2+X_1+X_2) (x_1+X_1+Y) (x_2+X_2+Y)}
    \nn\\&\quad
    +\frac{2\vep\ Y \  \d x_1\wedge \d x_2\wedge \d X_1}{x_1 (x_1+x_2+X_1+X_2) (x_1+X_1+Y) (x_2+X_2+Y)}
    \nn\\&\quad
    +\frac{2\vep\ Y\ \d x_1\wedge \d x_2\wedge \d X_2}{x_2 (x_1+x_2+X_1+X_2) (x_1+X_1+Y) (x_2+X_2+Y)}
    \nn\\&\quad 
    + \text{kinematic 2-forms}
    \,.
\end{align}
Notice that the right hand side above only has simple poles! Using the formulae given in appendix \ref{app:intNum}, one finds the first row of the differential equation $\mat{A}_\vth$
\begin{align} \label{eq:dthProj}
    (A_{\vth})_{1j} = (C_{\vth}^{-1})_{jk}
    \la \cvphi_k \vert (\D+\D\log(u)\wedge) \ {\color{DarkOrchid}\vth_1} \ra
    \,.
\end{align}
The above formula is attractive since it only involves algebraic (residue) operations. We know where the intersection number localizes and don't have to go through the process of constructing IBP-forms. 

Before moving on to actually integrating our differential equation, we note that the ideas outlined here are applicable to the algorithm for computing the symbol in \cite{Chen:2023kgw}. 
By promoting the external derivative to a derivative on the total space, the two-step algorithm for computing the intersection number could potentially be reduced to one-step!

\subsection{Integrating the differential equations \label{sec:4ptIntegration}}

Having constructed the DEQ system for the two-site/four-point FRW correlator in subsections \ref{sec:4ptBad} ($\vth$-basis) and \ref{sec:4ptSimp} ($\vphi$-basis), we now turn our attention to solving them. 
We implement conventional techniques that have become widespread in the computation of Feynman integrals over the last decade \cite{Henn:2014qga, Abreu:2022mfk}. 
Differential equations of the form constructed in sections \ref{sec:4ptBad} and \ref{sec:4ptSimp} are known as Pfaffian systems whose solutions are given by the path-ordered exponential as expressed in equation (\ref{eq:DEQsolpoe}). 
For small $\vep$, as is the case for FRW cosmologies, such solutions can be expanded and truncated to any desired order in $\vep$. 
In this subsection, we will reproduce the well known de Sitter result corresponding to the $\mathcal{O}(\vep^0)$ term of the FRW wavefunction coefficient.
Then, for the first time, we derive an expression for the $\mathcal{O}(\vep)$ term of the two-site/four-point FRW wavefunction coefficient.  

We begin by reviewing how to integrate systems of coupled linear DEQs in the context of the two-site/four-point example. The physical integral $\vth_1$ can be obtained by directly integrating the $\mat{A}_\vth$ system associated to the canonical basis, for which $\vth_1$ is a basis element, or by integrating the $\mat{A}_\vphi$ system and using the fact that ${\color{DarkOrchid}\vth_1} = \vphi_2+\vphi_3-\vphi_4$. 

In either case, the DEQ is 
\begin{align}
    \nabla_\text{kin} \alpha_i = A_{\alpha;ij} \wedge \alpha_j~,
\end{align}
where $\alpha = \vth, \vphi$ denotes our choice of basis. 
In terms of actual integrals instead of differential forms, this DEQ becomes 
\begin{align}
    \d_\text{kin} \, \vec{g}_\alpha(z, \vep) = \mat{A}_\alpha \cdot \vec{g}_\alpha(z, \vep)~,
    \label{eq:4ptDEQcomnotn}
\end{align}
where $\vec{g}_{\alpha} = \int u\, \alpha$ is the vector of integrals associated to our choice of basis. Here, $z \in \{X_1, X_2, Y\}$ refers to the external scales for the tree-level two-site/four-point example at hand and $\d_\text{kin} = \sum_{\{z\}} \partial_{z} \d z = \partial_{X_1} \d X_1 + \partial_{X_2} \d X_2 + \partial_Y \d Y$ as before. The $4 \times 4$ DEQ matrix $\mat{A}_\alpha$ can be expanded as
\begin{align}
    \mat{A}_\alpha = \vep \sum_{\{z\}} \mat{\Omega}_z \d z = \vep \left(\mat{\Omega}_{X_1} \d X_1 + \mat{\Omega}_{X_2} \d X_2 + \mat{\Omega}_{Y} \d Y\right)~,
\end{align}
where the entries of the $\mat{\Omega}_z$ matrices are rational one-forms. The solution $\vec{g}_\alpha(z,\vep)$ is a Laurent series in $\vep$
\begin{align}
    \vec{g}_\alpha(z,\vep) = \sum_{n=-2}^{\infty} \vep^n\ \vec{f}_\alpha^{(n)}(z)~,
    \label{eq:4ptDEQsolseries}
\end{align}
where the minimal value of $n$ is determined by boundary conditions (in this case, it happens to be $-2$). 

For the two-site/four-point example, the boundary conditions are generated by taking the $Y\to0$ limit where the integrals can actually be performed and then expanded in $\vep$. We denote this boundary condition by $\vec{g}^*_\alpha$
\begin{align}
    \vec{g}_\alpha^{*}(X_1,X_2,\vep) 
    := \vec{g}_\alpha^{*}(X_1,X_2,0,\vep) 
    = \sum_{n=-2}^{\infty} \vep^n \, \vec{f}^{*(n)}_\alpha(X_1,X_2)~.
\end{align}
For the $\vth$-basis, the boundary condition integrals are straightforward to compute. On the other hand, the integral associated to $\vphi_1$ is divergent. However, due to the simplicity of the $\mat{A}_\vphi$ DEQ, it is easy to see that the function $a(\vep) (X_1+X_2)^{2\vep}$ is a solution for the first component of $\vec{g}_\vphi$. Expanding $a(\vep)$ as a series in $\vep$, each coefficient can be fixed by demanding that the vector of boundary conditions $\vec{g}_\vphi^*$ satisfies the DEQ. In particular, using $a(\vep) = - \frac{1}{2\vep^2} - \frac{\pi^2}{4} + \mathcal{O}(\vep)$ is enough to fix the physical integral to $\mathcal{O}(\vep)$.

Having generated the boundary conditions, we are now in a position to integrate the DEQ system at hand. Given our Laurent series ansatz (\ref{eq:4ptDEQsolseries}), the DEQ system takes the form
\begin{align}
     \partial_z \, \vec{g}_\alpha(z, \vep) = \mat{\Omega}_z \, \cdot \vec{g}_\alpha(z, \vep) \implies \begin{cases}
         \partial_z \vec{f}^{(-2)} = 0 \\
         \partial_z \vec{f}^{(a)} = \mat{\Omega}_z \cdot \vec{f}^{(a-1)} \quad \forall~ a > 2~.
     \end{cases}
\end{align}
Following conventional DEQ techniques, we can now solve order by order in the twist parameter $\vep$. 
Thus, we are led to the following solution at order $\vep = 0$ for the two-site/four-point correlator, i.e., the solution corresponding to the physical chamber denoted by ${\color{DarkOrchid}\vth_1}$ 
\begin{align}
    \psi_\text{2,dS}^{(0)} &= 
        \text{Li}_2\left(\frac{X_1-Y}{X_1+Y}\right)
        + \text{Li}_2\left(\frac{X_2-Y}{X_2+Y}\right)
        - \text{Li}_2\left(
            \frac{X_1-Y}{X_1+Y}
            \frac{X_2-Y}{X_2+Y}
        \right)
        -\frac{\pi^2}{6}~.
    \label{eq:4ptsoldS}
\end{align}
Up to a sign,\footnote{The sign difference arises from the arbitrariness in the choice of orientation when constructing a canonical form.} 
the above solution corresponds to the well-know two-site contribution to the wavefunction coefficient in a dS Universe ($\vep = 0$) 
\cite{ArkaniHamed2015,Arkani-Hamed:2018kmz,Hillman:2019wgh}.

In order to go beyond dS, we solve the differential equations arising at the next order in $\vep$ (only for the physical component ${\color{DarkOrchid}\vth_1}$) leading us to the following two-site/four-point contribution to the wavefunction coefficient in a FRW Universe in the region $X_1>X_2>Y$
\begin{align}
   \psi_\text{2,FRW}^{(0)} =& 
    - \text{Li}_3\left(\frac{Y+X_1}{Y-X_2}\right)
    + \text{Li}_3\left(-\frac{X_1-Y}{Y+X_2}\right)
    + 2 \text{Li}_3\left(\frac{X_1-Y}{X_1+X_2}\right)
    - 2 \text{Li}_3\left(\frac{Y+X_1}{X_1+X_2}\right)
    \nn\\&
    + 2 \log \left(X_1+X_2\right) \text{Li}_2\left(-\frac{Y+X_2}{X_1-Y}\right)
    - 2 \log \left(X_1+X_2\right) \text{Li}_2\left(\frac{Y-X_2}{Y+X_1}\right)
    \nn\\&
    + \frac{1}{6} \log ^3\left(X_2-Y\right)
    - \frac{1}{6} \log ^3\left(X_2+Y\right)
    - \frac{1}{2} \log \left(X_1+Y\right) \log ^2\left(X_2-Y\right)
    \nn\\&
    - \frac{1}{2} \log ^2\left(X_1+Y\right) \log \left(X_2-Y\right)
    - \log ^2\left(X_1+X_2\right) \log \left(X_2-Y\right)
       \nn\\&
    +\frac{1}{2} \log \left(X_1+Y\right) \log ^2\left(X_2+Y\right)
    + \log ^2\left(X_1+X_2\right) \log \left(X_2+Y\right)
    \nn\\&
    + \frac{1}{2} \log ^2\left(X_1+Y\right) \log \left(X_2+Y\right)
    + \log \left(X_1+X_2\right) \log ^2\left(X_1-Y\right)
    \nn\\&
    - \log \left(X_1+X_2\right) \log ^2\left(X_1+Y\right)
    + 2 \log \left(X_1+X_2\right) \log \left(X_1+Y\right) \log \left(X_2-Y\right)
    \nn\\&
    + \frac{1}{6} \pi ^2 \left(\log \left(X_2-Y\right)
    - \log \left(X_2+Y\right)\right)
    \nn\\&
    - 2 \log \left(X_1+X_2\right) \log \left(X_1-Y\right) \log \left(X_2+Y\right)~.
   \label{eq:4ptsolFRW}
\end{align}
It is worth noting a few important features of the above expression. 
First, the integral representation of $\psi_{2,\text{FRW}}^{(0)}$ is \emph{not} manifestly symmetric in $X_1$ and $X_2$. We have sacrificed this symmetry for the simple presentation above. A formula valid in the whole physical region that is manifestly symmetric has been included in the ancillary files. 
Second, the above expression has uniform transcendental weight three. This is expected since the FRW wavefunction is the integral of the dS wavefunction thereby increasing the transcendental weight from two to three. 
In the next section, we will derive the differential equation for the tree-level three-site/five-point and one-loop two-site/two-point FRW wavefunction coefficients. While we leave the integration of these differential equation for future work, the resulting $\mathcal{O}(\vep)$ terms will have transcendental weight four and three since the corresponding dS wavefunction coefficients are know to be weight three and two respectively. 
Interestingly, loop-diagrams corresponding to cosmological correlators are not more transcendental than their tree-level counter parts because the integration over the spatial loop momentum is not performed. 

\section{Further examples \label{sec:further examples}}

In this section, we provide two further examples illustrating how to construct cosmological correlators associated to FRW integrals and the differential equations they satisfy. 
In particular, we show that there are no obstructions to extending the techniques applied in this work to loop integrands.\footnote{To go beyond and explicitly perform the integral over the spatial loop momenta, one needs to account for the integration measure, which introduces new polynomials into the numerator of the integrals \cite{Benincasa:2024lxe} that can be naturally accommodated in the framework of relative twisted cohomology. While polynomial numerators don't pose any technical problems, numerators with square roots may need additional regularization: $\sqrt{\bullet} \to (\bullet)^{\frac{1}{2}+\delta}$ where $\delta$ is a regulator much like the dimensional regulator. We leave the technical details of the $L$-loop FRW problem to upcoming work.}
To illustrate this point, we construct the cohomologies for the one-loop two-site/two-point FRW integrand as well as their differential equations in section \ref{sec:1L2S}. Then, in section \ref{sec:5pt}, we construct the cohomologies and associated differential equations for the tree-level three-site/five-point FRW wavefunction coefficient. Here, we see a new feature due to the degeneration of boundaries that will be present in most higher-site/point examples and explain how to account for this degeneracy.

\subsection{The one-loop two-site FRW correlator \label{sec:1L2S}}

\begin{figure}
    \centering
    \includegraphics[align=c, width=.3\textwidth]{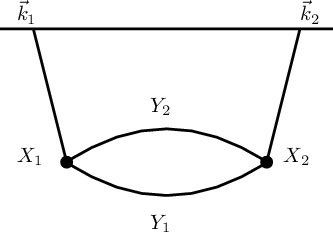}
    \qquad\qquad
    \includegraphics[align=c, width=.3\textwidth]{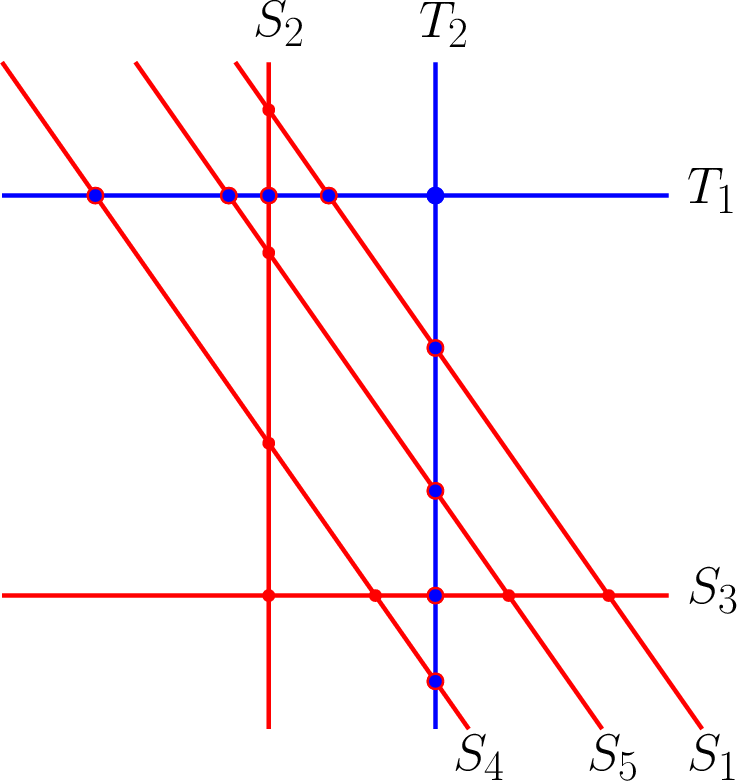}
    \caption{The Feynman diagram for the one-loop two-site/two-point wavefunction coefficient in $\lambda_3 \phi^3$ theory and the associated hyperplane arrangement. }
    \label{fig:1L2S}
\end{figure}

In $\lambda_3 \phi^3$ theory, the flat-space one-loop two-site/two-point wavefunction coefficient receives contributions from a single graph (see the left of figure \ref{fig:1L2S}) given by equation \eqref{eq:flatspace2-site1-loop}, which we quote here for convenience
\begin{align}
    \psi_{2,\text{flat}}^{(1)}
    &=
    \frac{2 \left(X_1+X_2+Y_1+Y_2\right)}{\left(X_1+X_2\right) \left(X_1+X_2+2 Y_1\right) \left(X_1+Y_1+Y_2\right) \left(X_2+Y_1+Y_2\right) \left(X_1+X_2+2 Y_2\right)}~.
\end{align}
The FRW uplift prescription \eqref{eq:FRWwavefunccoeff} yields the associated hyperplane arrangement for the FRW correlator (see the right of figure \ref{fig:1L2S}). 
As usual, the hyperplane arrangement is determined by the twisted coordinate hyperplanes 
\begin{align}
    T_i &= x_{i} \quad\text{for}\quad i=1,2~,
\end{align}
and the singular/boundary hyperplanes are defined by the linear divisors of $ \psi_{2,\text{flat}}^{(1)}\vert_{X_i \to x_i + X_i}$
\begin{align}
\begin{matrix*}[l]
    S_1 = x_1+x_2+X_1+X_2,
    &\quad&
    S_2 = x_1+X_1+Y_1+Y_2,
    \\
    S_3 = x_2+X_2+Y_1+Y_2,
    &\quad&
    S_4 = x_1+x_2+X_1+X_2+2 Y_1,
    \\
    S_5 = x_1+x_2+X_1+X_2+2 Y_2. 
\end{matrix*}
\end{align}
Note that this hyperplane arrangement is incredibly similar to the tree-level case and not any more difficult. 

Counting the bounded chambers in figure \ref{fig:1L2S} (right), we see that the FRW cohomology and its dual are 10-dimensional.
A convenient choice for the dual 1-boundary forms is
\begin{align} \label{eq:2S1L 1bd dual}
    \cvphi_1 = \vep\ \delta_1\left(\d\log \frac{T_1\vert_1}{T_2\vert_1}\right),
    \qquad
    \cvphi_2 = \vep\ \delta_4\left(\d\log \frac{T_1\vert_2}{T_2\vert_2}\right),
    \qquad
    \cvphi_3 = \vep\ \delta_5\left(\d\log \frac{T_1\vert_3}{T_2\vert_3}\right). 
\end{align}
A similarly convenient choice for the 2-boundary forms is
\begin{align}
\begin{matrix*}[l] \label{eq:2S1L 2bd dual}
    \cvphi_4 = \delta_{12}(1),
    \quad&
    \cvphi_5 = \delta_{13}(1),
    \quad&
    \cvphi_6 = \delta_{23}(1),
    \quad&
    \cvphi_7 = \delta_{24}(1),
    \\
    \cvphi_8 = \delta_{25}(1),
    &
    \cvphi_9 = \delta_{34}(1),
    &
    \cvphi_{10} = \delta_{35}(1).
    &
\end{matrix*}
\end{align}
A set of FRW forms dual to the 1-boundary forms \eqref{eq:2S1L 1bd dual} is 
\begin{align}
\begin{matrix*}[l]
    \vphi_1 = -\frac{1}{2}\ \d\log S_1 \wedge \d\log \frac{T_1}{T_2},
    &
    \vphi_2 = -\frac{1}{2}\ \d\log S_4 \wedge \d\log \frac{T_1}{T_2},
    \\
    \vphi_3 = -\frac{1}{2}\ \d\log S_5 \wedge \d\log \frac{T_1}{T_2},
    &
\end{matrix*}
\end{align}
while a set of FRW forms dual to \eqref{eq:2S1L 2bd dual} is 
\begin{align}
\begin{matrix*}[l]
    \vphi_4 = \d\log S_1 \wedge \d\log S_2,
    \qquad&
    \vphi_5 = \d\log S_1 \wedge \d\log S_3,
    \qquad &
    \vphi_6 = \d\log S_2 \wedge \d\log S_3,
    \\
    \vphi_7 = \d\log S_2 \wedge \d\log S_4,
    &
    \vphi_8 = \d\log S_2 \wedge \d\log S_5,
    &
    \vphi_9 = \d\log S_3 \wedge \d\log S_4,
    \\
    \vphi_{10} = \d\log S_3 \wedge \d\log S_5.
    & & 
\end{matrix*}
\end{align}
With these choices for our basis, it is easy to see that only the diagonal intersection numbers are non-trivial. Moreover, the normalization of $-\frac{1}{2}$ in \eqref{eq:2S1L 1bd dual} ensures that our choice of basis are orthonormal: each $T_i\vert_\bullet$ contributes a $\frac{1}{-\vep}$ to the intersection number while the point at infinity does not contribute since of the forms \eqref{eq:2S1L 1bd dual} are regular at infinity.

Computing the differential equation using the methods of section \ref{sec:4ptSimp} or \ref{sec:4ptNoIBP} yields 
{\small
\begin{align} \label{eq:2Site1LoopDEQ}
    \mat{A} = \vep \left(
\begin{array}{cccccccccc}
 2 a_1 & 0 & 0 & 0 & 0 & 0 & 0 & 0 & 0 & 0 \\
 0 & 2 a_2 & 0 & 0 & 0 & 0 & 0 & 0 & 0 & 0 \\
 0 & 0 & 2 a_3 & 0 & 0 & 0 & 0 & 0 & 0 & 0 \\
 a_5{-}a_4 & 0 & 0 & a_4{+}a_5 & 0 & 0 & 0 & 0 & 0 & 0 \\
 a_7{-}a_6 & 0 & 0 & 0 & a_6{+}a_7 & 0 & 0 & 0 & 0 & 0 \\
 0 & 0 & 0 & 0 & 0 & a_5{+}a_6 & 0 & 0 & 0 & 0 \\
 0 & a_8{-}a_5 & 0 & 0 & 0 & 0 & a_5{+}a_8 & 0 & 0 & 0 \\
 0 & 0 & a_{10}{-}a_5 & 0 & 0 & 0 & 0 & a_5{+}a_{10} & 0 & 0 \\
 0 & a_6{-}a_9 & 0 & 0 & 0 & 0 & 0 & 0 & a_6{+}a_9 & 0 \\
 0 & 0 & a_6{-}a_{11} & 0 & 0 & 0 & 0 & 0 & 0 & a_6{+}a_{11} \\
\end{array}
\right)~.
\end{align}
}
Here, one should read the $a_i$'s above as $\d\log(a_i)$ where the $a_i$ are the letters
\begin{align}
   a_{i=1,\dots,11} = \bigg\{&
   X_1+X_2,\ 
   X_1+X_2+2 Y_1,\ 
   X_1+X_2+2 Y_2,\ 
   X_2-Y_1-Y_2,\ 
   \nn\\&
   X_1+Y_1+Y_2,\ 
   X_2+Y_1+Y_2,\ 
   X_1-Y_1-Y_2,\ 
   X_2+Y_1-Y_2,\
   \nn\\&
   X_1+Y_1-Y_2,\ 
   X_2-Y_1+Y_2,\ 
   X_1-Y_1+Y_2
   \bigg\}.
\end{align}
The known de Sitter limit of this alphabet \cite{Hillman:2019wgh} is a subset of the above FRW alphabet with the new letters being $X_1+Y_1+Y_2$ and $X_2+Y_1+Y_2$. 

After integrating the above differential equation, the physical integral can be computed by taking the appropriate linear combination. In our chosen basis, 
\begin{align} \label{eq:2Site1LoopPhys}
    \psi_{2,\text{FRW}}^{(1)} 
    = \frac{1}{4 Y_1 Y_2}
    \int u \left(
        -\vphi_4
        +\vphi_5
        +\vphi_6
        -\vphi_7
        -\vphi_8
        +\vphi_9
        +\vphi_{10}
    \right).
\end{align}

While one can get the FRW IBPs algorithm of section \ref{sec:4ptBad} to finish in a reasonable amount of time on a laptop for this example, the dual IBPs of section \ref{sec:4ptSimp} or the intersection algorithm of section \ref{sec:4ptNoIBP} becomes noticeably faster here. 
The simplicity of dual IBPs is essential for the next example. There, we could not get a \emph{simple} implementation of the algorithm in section \ref{sec:4ptBad} to finish in a reasonable time unless supplemented with knowledge of the dual differential equation.

Our basis choice is motivated by the geometry of the dual 
cohomology and produces a sparse differential equation.
While this is desirable, there may be better choices. 
Since only the un-twisted hyperplanes appear in the physical FRW form, it would be interesting if one could find a basis that factorizes the differential equation \eqref{eq:2Site1LoopDEQ} into physical and un-physical sectors that can be solved separately.\footnote{Indeed, it looks like this had been achieved to some extent in \cite{ltalk}. While the FRW cohomology of the 3-site/5-point is 26-dimensional, in \cite{ltalk} they seem to only need 16 elements to get the DEQ for the physical integral.}
We leave such questions to future work. 


\subsection{The tree-level three-site FRW correlator \label{sec:5pt}}

\begin{figure}
    \centering
    \includegraphics[width=.3\textwidth]{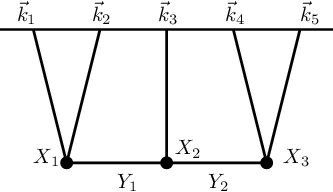}
    \caption{The Feynman diagram for the three-site/five-point FRW wavefunction coefficient in $\lambda_3 \phi^3$-theory.}
    \label{fig:3site}
\end{figure}

\begin{figure}
    \centering
    \includegraphics[width=.4\textwidth]{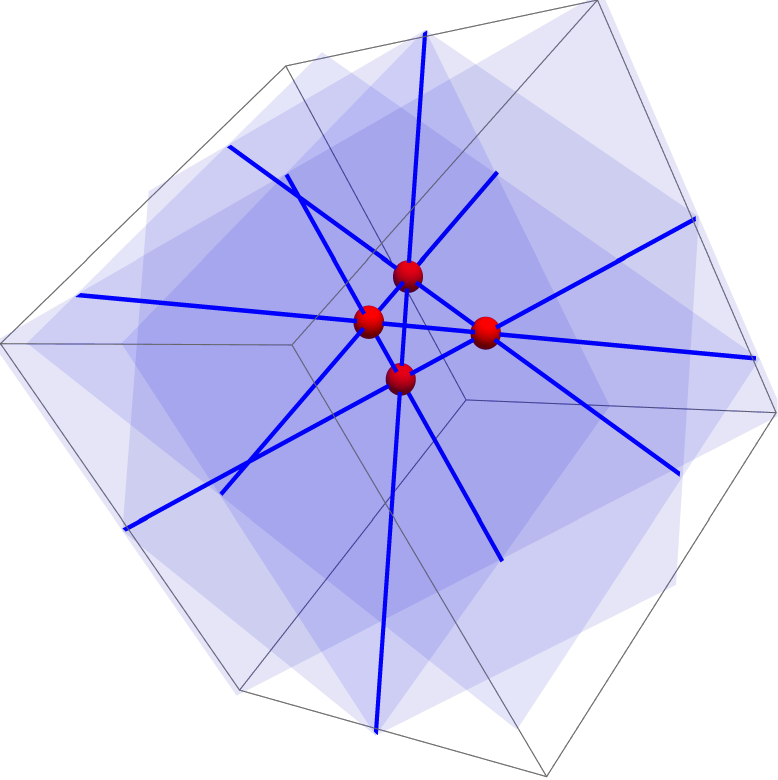}
    \qquad
    \includegraphics[width=.4\textwidth]{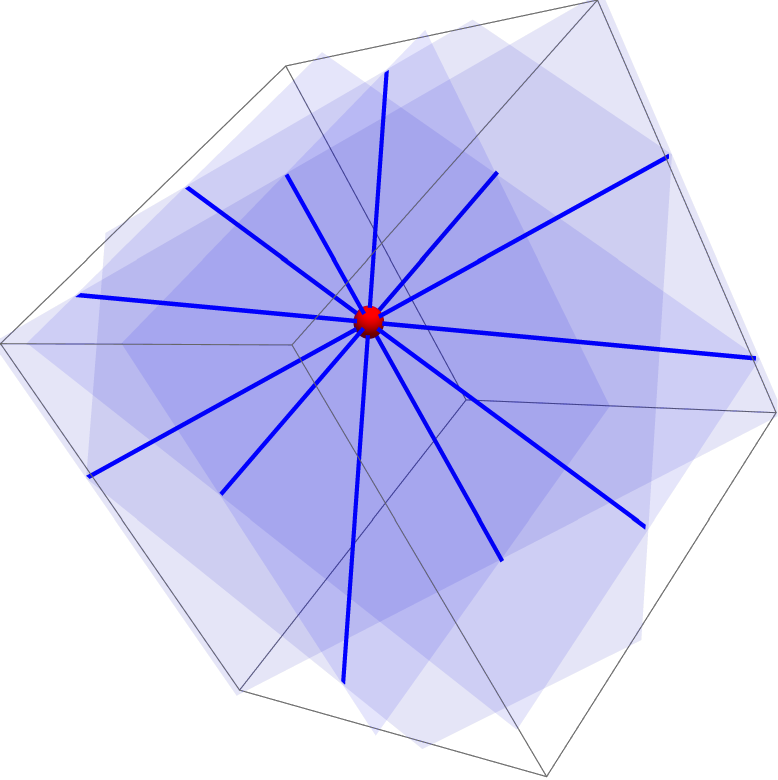}
    \caption{Plots of the planes $S_{i=1,4,5,6}$. The dark blue lines denote the intersection of these planes and the red dots mark where these lines intersect. The left plot corresponds to the unphysical but generic hyperplane arrangement with $Y_3\neq0$. The right plot is the degeneration of the left plot in the $Y_3\to0$ limit. The bounded chamber with the red dots as vertices in the right plot vanishes in the $Y_3\to0$ limit. This degeneration corresponds to the drop in dimension of the cohomology group. 
    }
    \label{fig:ArrangementDegen}
\end{figure}

We now turn our attention to the tree-level three-site/five-point FRW wavefunction coefficient where we will find new features not seen at the level of the two-site/four-point example but are expected to be present for most higher-site diagrams.
Substituting equation \eqref{eq:flatspace3-site} into (\ref{eq:FRWwavefunccoeff}), one obtains
\begin{align} \label{eq:3siteIntegral}
    \psi^{(0)}_{3,\text{FRW}} = 
    \int_0^{\infty} \left(\bigwedge_{i=1}^3 \d x_i\right)
    \left.
        \left(\prod_{a=1}^3 T_a^\vep\right)
        \psi^{(0)}_{3,\textrm{flat}}
    \right\vert_{X_i \to x_i + X_i}~,
\end{align}
where we duplicate the flat space three-site wavefunction coefficient \eqref{eq:flatspace3-site} below for ease of reading
\begin{align}
\psi_{3,\textrm{flat}}^{(0)} 
    &= \scalebox{1.25}{$
    \frac{X_1{+}Y_1{+}2X_2{+}Y_2{+}X_3}{(X_1{+}X_2{+}X_3) (X_1{+}Y_1) (X_3{+}Y_2)  (X_2{+}Y_1{+}Y_2) (X_1{+}X_2{+}Y_2) (X_2{+}X_3{+}Y_1)}$
    }.
\end{align}
The corresponding hyperplane arrangement is determined by the twisted coordinate hyperplanes
\begin{align}
    T_i = x_{i} \quad\text{for}\quad i=1,2,3,
\end{align}
and the linear denominators of $\psi^{(0)}_{3,\text{flat}}\vert_{X_i \to x_i + X_i}$
\begin{align}
\begin{matrix*}[l]
    S_1 = x_1 + x_2 + x_3 + X_1 +X_2 + X_3 {\color{lightgray} + Y_3}~,
    & \quad & 
    S_2 = x_1 + X_1 + Y_1~,
    \\
    S_3 = x_3 + X_3 + Y_2~,
    & &
    S_4 = x_2 + X_2 + Y_1 + Y_2~,
    \\
    S_5 = x_1 + x_2 + X_1 + X_2 + Y_2~,
    & &
    S_6 = x_2 + x_3 + X_2 + X_3 + Y_1~.
\end{matrix*}
\end{align}
Note that we have added an extra factor of $Y_3$ into $S_1$ that is not present in the physical arrangement. This has been done for purely pedagogical reasons since the physical arrangement ($Y_3=0$) is non-generic -- there are 4-planes crossing at a single point (see figure \ref{fig:ArrangementDegen}). 
Adding a non-zero $Y_3$ will help us understand how to think about non-generic arrangements. 
However, in the end, one can directly set $Y_3=0$ at the beginning of the calculation as we will explain later.

\begin{figure}
    \centering
    \includegraphics[width=\textwidth]{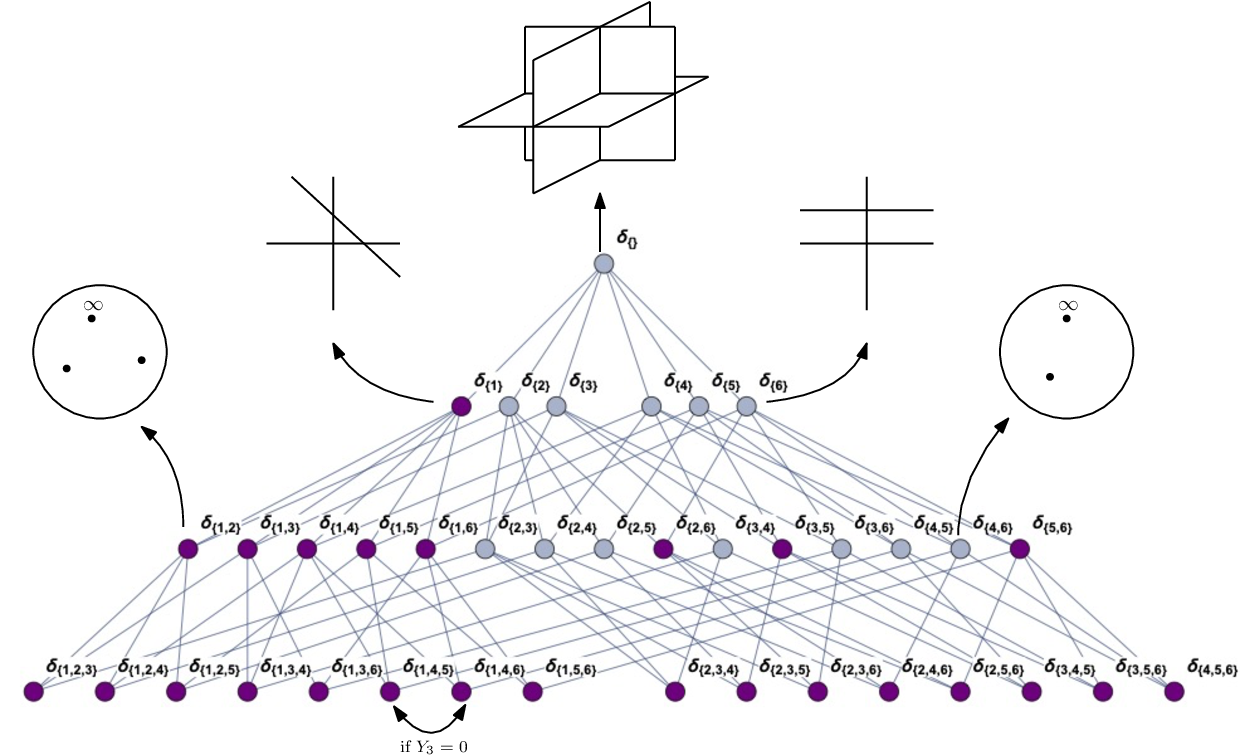}
    \caption{Boundary stratification of the relative twisted cohomology associated to the three-site/five-point hyperplane arrangement. The purple vertices are boundaries with non-trivial cohomology while all the blue vertices are boundaries with trivial cohomology. Notice that there are some missing 3-boundaries, $\delta_{126}, \delta_{135}, \delta_{245}, \delta_{346}$, 
    since $S_1 \cap S_2 \cap S_6 = S_1 \cap S_3 \cap S_5 = S_2 \cap S_4 \cap S_5 = S_3 \cap S_4 \cap S_6 = \emptyset$.
    }
    \label{fig:5ptBds}
\end{figure}

For the generic arrangement ($Y_3\neq0$), the basis size or dimension of the cohomology group is 26. The basis size is obtained by adding up the dimension of each twisted cohomology group on each of the non-maximal codimension boundaries.\footnote{Recall that the dimension of the top-dimensional twisted cohomology group is given by the Euler characteristic of the underlying manifold or equivalently by counting the critical points of the dual connection $\com$ (see footnote \ref{foot:critPoints}) or by counting the number of bounded chambers formed by the twisted hyperplanes.} 
Additionally, (for generic arrangements) each maximal codimension boundary generates a cohomology class/basis element. The boundary stratification for the three-site/five-point arrangement is presented in figure \ref{fig:5ptBds}. Each node denotes the twisted cohomology on the given boundary. The blue nodes represent boundaries with trivial cohomology (zero dimensional) while the purple nodes represent boundaries with non-trivial cohomology. Example arrangements of the twisted hyperplanes on some boundaries are illustrated in figure \ref{fig:5ptBds}. Since the twisted hyperplanes do not form bounded chambers in the bulk space $\delta_{\{\}}$ only the cohomologies on the boundaries are non-trivial. 
The 1-boundary $\delta_6$ is also pictured in figure \ref{fig:5ptBds} and is clearly trivial since there are no bounded chambers. 
On the other hand, the cohomology on $\delta_{1}$ is 1-dimensional since the twisted hyperplanes form a single bounded chamber. 
For the cohomologies on 2-boundaries, the dimension is determined by the number of twisted singularities minus two. 
Thus, the dimension on boundary $\delta_{12}$ is 1-dimensional while the cohomology on the boundary $\delta_{46}$ is trivial. For the generic arrangement, the untwisted 3-boundaries are all distinct and 1-dimensional. 

For the physical ($Y_3=0$) arrangement, the basis size decreases to 25 because the four hyperplanes $S_1, S_4, S_5$ and $S_6$ all intersect at a point (see figure \ref{fig:ArrangementDegen}). Sending $Y_3\to0$ shrinks the bounded chamber pictured in the left of figure \ref{fig:ArrangementDegen} to a point decreasing the basis size. Since we expect that there will be many more instances of such degeneracies for higher site FRW integrals, we provide a careful treatment of how the generic arrangement degenerates to the physical arrangement below.\footnote{For the four-site graph, there are two topologically distinct Feynman diagrams: a line and a three-pointed star. For example, the hyperplane arrangement corresponding to the \emph{generic} three-point star graph consists of 11 boundary hyperplanes in generic position and the 3 twisted coordinate hyperplanes as usual. We find that this generic arrangement of the star graph has a total of 377 elements. The degenerate/physical arrangement has 12 lines where 4 of the boundary hyperplanes intersect, 18 points where 5 of the boundary hyperplanes intersect and 1 point where 8 of the boundary hyperplanes intersect. This is expected to reduce the number of independent elements by a considerable amount.}
Again, we emphasize that one does not need to introduce additional parameters to make the arrangement generic as long as one knows how to consistently treat the non-generic/degenerate case directly. 

Below, we give a natural basis associated to the generic arrangement where the element that drops out when passing to the physical arrangement has been colored in gray.\footnote{While we have chosen to eliminate the forms associated with the intersection $S_1\cap S_4 \cap S_5$, we could have chosen to eliminate the form associated to the intersections $S_1\cap S_5 \cap S_6$, $S_1\cap S_4 \cap S_6$ or $S_4\cap S_5 \cap S_6$ instead.} The only non-trivial 1-boundary cohomology occurs on the boundary $\delta_{1}$ and is generated by the single form
\begin{align} \label{eq:1bd dual basis}
    \cvphi_1 &= \vep^2\ \delta_{1}\left(
        \d\log\left(\frac{T_1\vert_1}{T_3\vert_1}\right)
        \wedge
        \d\log\left(\frac{T_3\vert_1}{T_2\vert_1}\right)
    \right).
\end{align}
This is nothing but the canonical form associated to the bounded chamber on the boundary $\delta_{1}$. We have also normalized by $\vep^2$ since we expect the intersection of 2-boundary forms to be proportional to $1/\vep^2$. 
The 2-boundary cohomologies are generated by the following 9 forms
\begin{align} \label{eq:2bd dual basis}
    \begin{array}{ccc}
        \cvphi_2 = \vep\ \delta_{12}\left(
            \d\log\frac{T_2\vert_{12}}{T_3\vert_{12}}
        \right), \,
        & 
        \cvphi_3 = \vep\ \delta_{13}\left(
            \d\log\frac{T_1\vert_{13}}{T_2\vert_{13}}
        \right), \,
        & 
        \cvphi_4 = \vep\ \delta_{14}\left(
            \d\log\frac{T_1\vert_{14}}{T_3\vert_{14}}
        \right), \,
        \\
        \cvphi_5 = \vep\ \delta_{15}\left(
            \d\log\frac{T_1\vert_{15}}{T_2\vert_{15}}
        \right), \,
        &
        \cvphi_6 = \vep\ \delta_{16}\left(
            \d\log\frac{T_2\vert_{16}}{T_3\vert_{16}}
        \right), \,
        &
        \cvphi_7 = \vep\ \delta_{26}\left(
            \d\log\frac{T_2\vert_{26}}{T_3\vert_{26}}
        \right), \,
        \\
        \cvphi_8 = \vep\ \delta_{35}\left(
            \d\log\frac{T_1\vert_{35}}{T_2\vert_{35}}
        \right), \, 
        &
        \cvphi_9 = \vep\ \delta_{56}\left(
            \d\log\frac{T_1\vert_{56}}{T_2\vert_{56}}
        \right), \,
        &
        \cvphi_{10} = \vep\ \delta_{56}\bigg(
            \d\log T_3\vert_{56}
        \bigg).
    \end{array}
\end{align}
Note that the cohomology on all 2-boundaries is 1-dimensional except for the $\delta_{56}$ boundary, which is 2-dimensional. The 2-boundary forms are normalized by a single power of $\vep$ since we expect the 2-boundary intersection to be proportional to $1/\vep$. 
Finally, the 3-boundary cohomologies are generated by the following 16 forms
\begin{align}\label{eq:3bd dual basis}
    \begin{array}{cccc}
         \cvphi_{11} = \delta_{123}\left(1\right)
         , \quad
         &  
         \cvphi_{12} = \delta_{124}\left(1\right)
         , \quad 
         &
         \cvphi_{13} = \delta_{125}\left(1\right)
         , \quad 
         &
         \cvphi_{14} = \delta_{134}\left(1\right)
         , \quad 
         \\
         \cvphi_{15} = \delta_{136}\left(1\right)
         , \quad
         &  
         {\color{lightgray}
         \cvphi_{16} = \delta_{145}\left(1\right)
         , \quad 
         }
         &
         \cvphi_{17} = \delta_{146}\left(1\right)
         , \quad 
         &
         \cvphi_{18} = \delta_{156}\left(1\right)
         , \quad 
         \\
         \cvphi_{19} = \delta_{234}\left(1\right)
         , \quad
         &  
         \cvphi_{20} = \delta_{235}\left(1\right)
         , \quad 
         &
         \cvphi_{21} = \delta_{236}\left(1\right)
         , \quad 
         &
         \cvphi_{22} = \delta_{246}\left(1\right)
         , \quad 
         \\
         \cvphi_{23} = \delta_{256}\left(1\right)
         , \quad
         &  
         \cvphi_{24} = \delta_{345}\left(1\right)
         , \quad 
         &
         \cvphi_{25} = \delta_{356}\left(1\right)
         , \quad 
         &
         \cvphi_{26} = \delta_{456}\left(1\right)
         . \quad 
    \end{array}
\end{align}
Notice that there are some missing 3-boundaries ($\delta_{126}, \delta_{135}, \delta_{245}, \delta_{346}$) since some triples of the relative surfaces do not intersect at a point: $S_1 \cap S_2 \cap S_6 = S_1 \cap S_3 \cap S_5 = S_2 \cap S_4 \cap S_5 = S_3 \cap S_4 \cap S_6 = \emptyset$.

Next we provide a basis of $\d\log$-forms orthonormal to the above dual basis. A $\d\log$-form dual to the 1-boundary form \eqref{eq:1bd dual basis} is 
\begin{align} \label{eq:1bd basis}
    \vphi_1 &= \frac{1}{3}\ \d\log S_1 \wedge 
        \d\log\left(\frac{T_1}{T_3}\right)
        \wedge
        \d\log\left(\frac{T_3}{T_2}\right).
\end{align}
The normalization of $1/3$ ensures that the intersection of \eqref{eq:1bd basis} with \eqref{eq:1bd dual basis} is unity. 
It is a simple exercise to see that this is indeed the correct normalization.
First, it is easy to see that $\vphi_1$ only has a non-trivial overlap with $\cvphi_1$.
Their intersection $\la \cvphi_1 \vert \vphi_1 \ra$ is proportional to the self-intersection of the canonical form corresponding to the chamber bounded on $X \cap S_1$. 
Thus, there are three-intersection points contributing to the intersection number. Since each $T_i$ has the same twist, we expect each intersection point to contribute with the same factor. Hence, the normalization of $1/3$.

A set of $\d\log$-forms orthonormal to the 2-boundary forms \eqref{eq:2bd dual basis} is
\begin{align} \label{eq:2bd basis}
\begin{array}{cc}
    \vphi_2 = -\frac{1}{2}\
        \d\log S_1 \wedge \d\log S_2 \wedge 
        \d\log\frac{T_2}{T_3},
    \qquad &
    \vphi_3 = -\frac{1}{2}\
        \d\log S_1 \wedge \d\log S_3 \wedge
        \d\log\frac{T_1}{T_2},
    \\
    \vphi_4 = -\frac{1}{2}\
        \d\log S_1 \wedge \d\log S_4 \wedge 
        \d\log\frac{T_1}{T_3},
    \qquad &
    \vphi_5 = -\frac{1}{2}\
        \d\log S_1 \wedge \d\log S_5 \wedge
        \d\log\frac{T_1}{T_2},
    \\
    \vphi_6 = -\frac{1}{2}\
        \d\log S_1 \wedge \d\log S_6 \wedge
        \d\log\frac{T_2}{T_3},
    \qquad &
    \vphi_7 = -\frac{1}{2}\
        \d\log S_2 \wedge \d\log S_6 \wedge
        \d\log\frac{T_2}{T_3},
    \\
    \vphi_8 = -\frac{1}{2}\
        \d\log S_3 \wedge \d\log S_5 \wedge
        \d\log\frac{T_1}{T_2},
    \qquad &
    \vphi_9 = -\frac{1}{2}\
        \d\log S_5 \wedge \d\log S_6 \wedge 
        \d\log\frac{T_1}{T_2},
    \\
    \vphi_{10} = -\frac{3}{2}\
        \d\log S_5 \wedge \d\log S_6 \wedge 
        \d\log T_3 .
    \qquad &
\end{array}
\end{align}
Following a similar argument as that below \eqref{eq:1bd dual basis}, the normalization's of these forms are simple to fix. The relative sign difference between the normalization of \eqref{eq:2bd basis} and \eqref{eq:1bd basis} comes from the fact that the dual twist has exponent $-\vep$ and for each boundary we lose a power of $-\vep$ in the intersection number. 
Furthermore, the reason why the normalization of $\vphi_{10}$ differs from the rest is because it is the only form where the point at infinity (on $X \cap S_5 \cap S_6$) contributes to its self-intersection. 

Lastly, a set of $\d\log$ forms orthonormal to the 3-boundary forms \eqref{eq:3bd dual basis} is
\begin{align} \label{eq:3bd basis}
\begin{matrix*}[l]
    \vphi_{11} =\d\log S_1 \wedge \d\log S_2 \wedge \d\log S_3 ,
    \qquad\quad\,
    &
    \vphi_{12} = \d\log S_1 \wedge \d\log S_2 \wedge \d\log S_4 , 
    \\
    \vphi_{13} = \d\log S_1 \wedge \d\log S_2 \wedge \d\log S_5 ,
    &
    \vphi_{14} = \d\log S_1 \wedge \d\log S_3 \wedge \d\log S_4 ,
    \\
    \vphi_{15} = \d\log S_1 \wedge \d\log S_3 \wedge \d\log S_6 ,
    &
    {\color{lightgray}
    \vphi_{16} = \d\log S_1 \wedge \d\log S_4 \wedge \d\log S_5 ,
    }
    \\
    \vphi_{17} = \d\log S_1 \wedge \d\log S_4 \wedge \d\log S_6 ,
    &
    \vphi_{18} = \d\log S_1 \wedge \d\log S_5 \wedge \d\log S_6 - \vphi_{10},
    \\
    \vphi_{19} = \d\log S_2 \wedge \d\log S_3 \wedge \d\log S_4 ,
    &
    \vphi_{20} =\d\log S_2 \wedge \d\log S_3 \wedge \d\log S_5 ,
    \\
    \vphi_{21} = \d\log S_2 \wedge \d\log S_3 \wedge \d\log S_6 ,
    &
    \vphi_{22} = \d\log S_2 \wedge \d\log S_4 \wedge \d\log S_6 ,
    \\
    \vphi_{23} = \d\log S_2 \wedge \d\log S_5 \wedge \d\log S_6 - \vphi_{10},
    &
    \vphi_{24} = \d\log S_3 \wedge \d\log S_4 \wedge \d\log S_5 ,
    \\
    \vphi_{25} = \d\log S_3 \wedge \d\log S_5 \wedge \d\log S_6 - \vphi_{10},
    &
    \vphi_{26} = \d\log S_4 \wedge \d\log S_5 \wedge \d\log S_6 {-} \vphi_{10}.
\end{matrix*}
\end{align}
To have a orthogonal basis, we have to subtract $\vphi_{10}$ from forms with non-trivial $S_5 \cap S_6$ residues since the intersections $\la \cvphi_{10} \vert \d\log S_\bullet \wedge \d\log S_5 \wedge \d\log S_6 \ra \sim \la \d\log T_3\vert_{56} \vert \d\log S_\bullet \ra \neq 0$ are non-vanishing. This is because both forms share a singularity at infinity on $X \cap S_5 \cap S_6$ and why it is best to choose $\d\log$-forms built out of ratios of the twisted hyperplanes when constructing a dual basis.

\begin{figure}
    \centering
    \includegraphics[width=.3\textwidth]{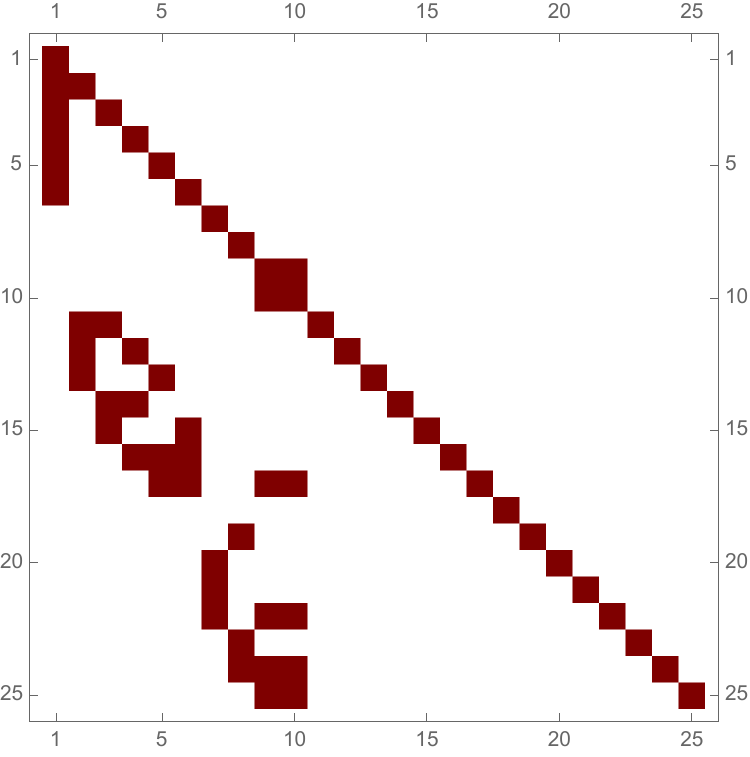}
    \caption{Matrix plot of the connection $\mat{A}_\text{phys} = -\mat{\c{A}}_\text{phys}^\top$. Each colored square on this $25 \times 25$ grid represents a non-zero element of $\mat{A}_\text{phys}$. Note that this DEQ is very sparse!}
    \label{fig:AMatrixPlot}
\end{figure}

Computing the dual differential equation using the method outlined in section \ref{sec:4ptSimp} or \ref{sec:4ptNoIBP} yields 
\begin{align}
    \bs{\c{\nabla}} \bs{\cvphi} &= \vep\ \mat{\c{A}} \cdot \bs{\cvphi},
    \\ 
    \mat{\c{A}} &= \sum \mat{\c{A}}_i\  \d\log a_i, 
\end{align}
where the $a_i$ are the letters in table \ref{tab:GenAlphabet} and the $\mat{\c{A}}_i$ are $26\times26$ $\mathbb{Q}$-valued matrices. The exact form of these coefficient matrices can be found in the ancillary file (see figure \ref{fig:AMatrixPlot} for the general structure or equation \ref{eq:3siteAphys} for the explicit result). 
Of course, since $\la \cvphi_a \vert \vphi_b \ra = \delta_{ab}$, we also have $\nabla \vphi_a = \vep\ A_{ab} \wedge \vphi_b$ where $\mat{A} = - (\c{\mat{A}})^\top$.

\begin{table}[]
    \centering
    \begin{tabular}{cccc}
        $X_1+X_2+X_3 {\color{lightgray}+Y_3},$ 
        &  
        $X_1+Y_1,$
        &
        $X_2+X_3-Y_1 {\color{lightgray}+Y_3},$
        &
        $X_3+Y_2,$
        \\
        $X_1+X_2-Y_2 {\color{lightgray}+Y_3},$
        & 
        {\color{BrickRed}
        $X_2+Y_1+Y_2,$
        }
        &
        $X_1+X_3-Y_1-Y_2 {\color{lightgray}+Y_3},$
        &
        $X_1+X_2+Y_2,$
        \\
        ${\color{NavyBlue} X_3-Y_2} {\color{lightgray}+Y_3},$
        &
        $X_2+X_3+Y_1,$
        &        
        ${\color{ForestGreen}X_1-Y_1} {\color{lightgray}+Y_3},$
        &
        $X_2-Y_1-Y_2 {\color{lightgray}+Y_3},$
        \\
        $X_3-2 Y_1-Y_2 {\color{lightgray}+Y_3},$
        &
        $X_2-Y_1+Y_2,$
        &
        $X_1-Y_1-2 Y_2 {\color{lightgray}+Y_3},$
        &
        $X_2+Y_1-Y_2,$
        \\
        {\color{ForestGreen}
        $ X_1-Y_1,$
        }
        &
        {\color{NavyBlue}
        $X_3-Y_2,$
        }
        &
        ${\color{BrickRed}X_2+Y_1+Y_2} {\color{lightgray}-Y_3},$
        &
        $X_3+2 Y_1-Y_2,$
        \\
        &
        $X_1-Y_1+2 Y_2,$
        &
        $X_1-X_3-Y_1+Y_2$
        &
    \end{tabular}
    \caption{The alphabet for the differential equation $a_i$. For the generic arrangement, there are a total of 22 different letters. None of these letters are singular in the $Y_3\to0$ limit. The colored pairs of letters correspond to letters that are to be identified in the $Y_3\to0$ limit. Thus, the physical DEQ has only 19 letters. The symbol alphabet of the de Sitter three-site/five-point integral forms a subset of the physical 19 letters as expected \cite{ArkaniHamed2017,Hillman:2019wgh}.
    }
    \label{tab:GenAlphabet}
\end{table}

Staring from the generic arrangement we will illustrate how to degenerate $\mat{\c{A}}$ to the physical arrangement. This will suggest a way to treat the physical case directly without deforming the physical arrangement. Both methods have been checked to yield the same resulting differential equation.

In the limit $Y_3\to0$, the boundaries (or equivalently residues) $\delta_{145},\delta_{146},\delta_{156}$ and $\delta_{456}$ are not independent since the following intersection matrix does not have full rank
\begin{align}
    \la \cvphi_a \vert \vphi_b \rangle
    = \begin{pmatrix}
        \,1 & 1 & 0 & 0 
        \\
        \,1 & 1 & 0 & 0
        \\
        -1 & 0 & 1 & 0 
        \\
        \,1 & 0 & 0 & 1
    \end{pmatrix}
    \quad 
    \text{where}
    \quad
    a,b \in \{16,17,18,26\}.
\end{align}
Choosing to eliminate $\vphi_{16}$ and $\cvphi_{16}$, yields 
\begin{align}
    \label{eq:DegenEquivMinus}
    &\cvphi_{16} = \cvphi_{17}
    \implies \delta_{145} = \delta_{146},
    \\
    \label{eq:DegenEquivPlus}
    &\vphi_{16} = \vphi_{17} - \vphi_{18} + \vphi_{26},
\end{align}
The second equation is an equivalence among $\d\log$-forms that can be checked directly. On the other hand, the first equation tells us that the residue operators $\res_{145}$ and $\res_{156}$ are equivalent and that we should identify $\delta_{145}$ and $\delta_{156}$.  Since all the $S_i \cap S_j \cap S_k$ are the same for $i,j,k\in\{1,4,5,6\}$ one might wonder why we do not identify all the $\delta_{ijk}$ with $i,j,k\in\{1,4,5,6\}$. Loosely speaking, the different $\delta_{ijk}$ with $i,j,k\in\{1,4,5,6\}$ represent the ``directions'' that one can approach the degenerate boundary. Such degenerate configurations often need careful treatment by blow-ups to resolve. Thankfully, the hyperplane arrangements associated to cosmological correlators 
are sufficiently simple that we can get by \emph{without} introducing additional mathematical formalism -- one simply finds the independent residues.

The physical DEQ 
\begin{align}
    \bs{\cvphi}_\text{phys} 
    &= ( \cvphi_1, \dots, \cvphi_{15}, \cvphi_{17}, \dots, \cvphi_{26} )^\top ~,
    \\
    \bs{\c{\nabla}} \bs{\cvphi}_\text{phys} 
    &= \vep\ \mat{\c{A}}_\text{phys} \cdot \bs{\cvphi}_\text{phys}~,
\end{align}
is obtained from $\mat{\c{A}}$ by eliminating the $16^\text{th}$ row and adding the $16^\text{th}$ column to the $17^\text{th}$ column as instructed by equation \eqref{eq:DegenEquivMinus}. For the explicit form of $\mat{A}_\text{phys}$, see the ancillary files or equation \eqref{eq:3siteAphys}. Figure \ref{fig:AMatrixPlot} also provides a rough sketch of the structure of $\mat{A}_\text{phys}$ without using up a whole page. Once again, a natural basis of the dual cohomology has lead to a remarkably sparse differential equation! 
As a consistency check, we note that the symbol alphabet for the dS wavefunction coefficient is a subset of our FRW symbol alphabet \eqref{eq:3siteAlpha}, as expected \cite{ArkaniHamed2017,Hillman:2019wgh}.

Equivalently, one could compute the physical DEQ without introducing $Y_3$ by making the identification $\delta_{145}=\delta_{156}$ at the start of the calculation. To avoid missing potential boundary elements of the dual cohomology, we recommend that one assumes that all boundaries are independent and then check for relations among them. 
Complications from non-generic arrangements are unavoidable and present in both the dual and FRW cohomologies. 

Once the $\mat{A}_\text{phys}$ system has been integrated, the three-site/five-point FRW integral \eqref{eq:3siteIntegral} can be recovered from the decomposition of its integrand
\begin{align}
    \psi^{(0)}_{3,\text{FRW}} 
    = \frac{1}{4Y_1Y_2} \int u\
    \left(
        \vphi_{11}
        {-}\vphi_{13}
        {+}\vphi_{15}
        {-}\vphi_{18}
        {+}\vphi_{19}
        {-}\vphi_{20}
        {-}\vphi_{21}
        {+}\vphi_{22}
        {-}\vphi_{24}
        {+}\vphi_{26}
    \right).
\end{align}
Since the dS wavefunction coefficient $\psi^{(0)}_{3,\text{dS}}$ has transcendental weight three, the leading FRW ($\mathcal{O}(\vep)$) term must have transcendental weight four.

\section{Discussions and future directions \label{sec:conclusion}}

The main goal of this work is to elucidate how (relative) twisted cohomology and intersection theory can be used to study the analytic structure of FRW integrals. 
In particular, we construct the differential equations obeyed by families of FRW integrals. 
The techniques presented here provide a geometric framework for understanding Euler/generalized hypergeometric integrals arising from hyperplane arrangements.

Families of FRW integrals form a particularly simple subset of such integrals where not all singularities are twisted. The un-twisted singularities correspond to boundaries in the associated dual cohomology leading to a simple geometric picture where dual forms are supported only on the boundaries. 
By working in the dual space, we gain both conceptual and computational advantages. In particular, integration-by-parts is much simpler in the dual cohomology since all forms are restricted to boundaries. 

After defining the proper dual cohomology, we can use the intersection number -- an inner product on the vector space of FRW integrals -- to decompose any FRW integral into a chosen basis. In particular, the intersection number can be used to derive differential equations for a family of FRW integrals bypassing traditional integration-by-parts algorithms. 
Intersection numbers have several advantages over integration-by-parts algorithms. 
Most importantly, the intersection number is an algebraic procedure given by a sequence of residues localizing onto the maximal codimension intersection points of the FRW hyperplane arrangement. 
The use of the intersection number for studying FRW integrals is efficient and practical since the intersection number of $\d\log$-forms is easy to compute. 

For various examples (tree-level two-site/four-point and three-site/five-point integrals as well as the one-loop two-site/two-point integral), we have constructed explicit representatives for a basis of the associated (relative) twisted cohomology. Then, using dual integration-by-parts identities or intersection numbers, we derived the corresponding differential equations. 
For the tree-level two-site/four-point example, we have explicitly integrated the differential equation to $\mathcal{O}(\vep)$ and obtained the first correction (\ref{eq:4ptsolFRW}) to the known de Sitter result (\ref{eq:4ptsoldS}). 

As already mentioned a few times throughout the text, finding/choosing a ``good'' basis is often artisanal. 
Motivated by the geometry of the dual cohomology, our choices for a good basis yielded simple intersection matrices and sparse differential equations. 
Yet, from \cite{ltalk} it seems that one may be able to choose a basis where the differential equation decomposes into physical and un-physical sectors that can be integrated separately. 
It would be interesting to further study the physical and mathematical principles behind this basis choice and subsequent factorization of the differential equations.\footnote{A preliminary analysis reveals that the dual forms in the physical sector correspond to unitarity cuts of the physical integrand. The full analysis of this fact and the possibility of a canonical basis of physical integrals is left for upcoming work.}

It would also be interesting to understand the cohomology in the de Sitter limit $\vep\to 0$ where the twist disappears. 
Such an understanding would also be useful in the context of $\mathcal{N}=4$ SYM amplitudes where one could hope to understand soft and collinear singularities cohomologically. 
Similarly, it would be interesting to understand the un-twisted cohomology of the $m=1$ amplituhedron, which has been identified as the complex of bounded faces of a cyclic hyperplane arrangement in \cite{Karp:2016uax}.

Another interesting question is whether the tools of intersection theory can be utilised to help understand how the constraints of bulk unitarity are encoded in cosmological correlators in a general FRW Universe (with a boundary). It has been recently pointed out in \cite{Goodhew:2020hob,Goodhew:2021oqg,Melville:2021lst,Baumann:2021fxj} that unitarity imposes specific constraints that inflationary correlators have to satisfy, closely resembling the optical theorem in flat space. These systematic cutting rules that cosmological correlators obey rely on very general properties of the Green's functions used to compute the wavefunction \cite{Baumann:2021fxj}. As a result, there lies no conceptual obstruction to generalizing these cutting rules to FRW wavefunction coefficients. However, with the Green's functions being rather complicated in a given FRW spacetime, such generalizations could be computationally cumbersome. Since intersection theory has also enhanced our understanding of $d$-dimensional generalized unitarity in multi-loop Feynman integrals \cite{Caron-Huot:2021iev,Caron-Huot:2021xqj}, it would be interesting to see if such techniques simplify some aspects of the cosmological cutting rules in a general FRW Universe, both at the level of trees and loops. 

In this paper, we considered correlation functions of a simplistic, toy-model FRW theory comprising solely of conformally coupled scalars with a cubic interaction. 
In such a scenario, our new result for the leading $\vep$ correction to the four-point function in dS (equation \ref{eq:4ptsolFRW}) could pave the way for understanding distinctive signatures during inflation and beyond. 
Like the analysis conducted in \cite{ArkaniHamed2015} where the four-point function in dS was utilised to understand distinctive signatures of primordial non-Gaussianities and also as a probe to predict new particles during inflation, the four-point function for $\vep \neq 0$ (equation \ref{eq:4ptsolFRW}) is perhaps a first step towards understanding physics in an inflationary regime described by a quasi-dS period or physics beyond the reheating surface. 
Moreover, correlators of conformally-coupled scalars act as seeds from which correlators of massless/massive fields with higher-spin exchanges can be deduced by acting with differential operators \cite{Baumann:2019oyu}. 
Thus, our analysis serves as an initial step towards  analysing physically relevant theories in FRW cosmologies and facilitating concrete cosmological predictions.

\section*{Acknowledgments}

AP would like to thank Sebastian Mizera for pointing out the relevance of hyperplane arrangements in the cosmological context, Nima Arkani-Hamed and Aaron Hillman for stimulating discussions, as well as the Institute for Advanced Study for its hospitality. 
SD and AP would also like to thank Heliudson de Oliveira Bernardo, Claudia Fevola and Henrik Munch for useful discussions. SD and AP are especially grateful to Anastasia Volovich and Marcus Spradlin for valuable feedback throughout the project and help in simplifying equation \eqref{eq:4ptsolFRW}. 
This work was supported in part by the US Department of Energy under contract DESC0010010 Task F (SD) and by Simons Investigator Award $\#376208$ (AP).

\appendix

\section{Simple formulae for the intersection number of logarithmic forms \label{app:intNum}}

Fortunately, FRW-forms and their duals are always logarithmic so that the needed intersection numbers are easy to compute and purely combinatorial. To avoid getting bogged down by unnecessary mathematical formalism, we provide simple formulas to compute logarithmic intersection numbers without proof. The interested reader is encouraged to consult \cite{matsumoto1998kforms, Mizera:2017rqa, Mizera:2019gea, Mizera:2019vvs, Frellesvig:2019uqt, Weinzierl:2020xyy, Chestnov:2022xsy, Caron-Huot:2021iev, Caron-Huot:2021xqj} and the references therein for more details. 

To compute the intersection number, one first takes care of any $\delta_I$'s via 
\begin{align}\label{eq:deltaInt}
    \bigg\la \delta_I(\cphi_I) \bigg\vert \vphi \bigg\ra
    =
    \bigg\la \cphi_I \bigg\vert \res_I\left[\frac{u}{u\vert_I}\vphi\right] \bigg\ra
    \underset{\text{if $\vphi$ $\d\log$}}{=} 
    \bigg\la \cphi_I \bigg\vert \res_I\left[\vphi\right] \bigg\ra
    \equiv 
    \la \cphi_I \vert \phi_I \ra~,
\end{align}
where $\phi_I$ is a short had for $\text{Res}_I[\vphi]$.  
Since FRW-forms have only simple poles on $S$, the ratio $u/u\vert_I$ is not needed and can be set to unity in the second equality above. 
It is also important to note that the residue is anti-symmetric with respect to the ordering of $I$. 
In our conventions, we define $\text{Res}_I$ by $\text{Res}_{I}[\bullet] = \text{Res}_{i_{|I|}} \cdots \text{Res}_{i_2} \text{Res}_{i_1}[\bullet]$.

After using up all the $\delta_J$'s, the remaining intersection number is an intersection number on the submanifold $X_I = X \cap S_I$.
We provide a simple formula for its computation that localizes to the maximal codimension intersection points of the twisted surfaces.
The algorithm presented here was first developed in \cite{matsumoto1998kforms} and we aim to provide a user friendly summary of this work.

The first step is to identify maximal intersection points of the twisted hyperplanes with themselves and the hyperplane at infinity ($H_\infty$): $\mathbb{CP}^{n-|I|}\setminus (T \cap S_I) \cup H_\infty$. Let us call this set $\text{Int}_I$. For each intersection point, choose \emph{normal} coordinates $\bs{z}^*$ such that the origin $\bs{0}$ corresponds to the intersection point. 
Then, the intersection number of logarithmic forms is given by 
\begin{align} \label{eq:CombInt}
    \bigg\la \cphi_I \bigg\vert \res_I\left[\vphi\right] \bigg\ra
    &=
        \sum_{\bs{z}^* \in \text{Int}_I} 
        \frac{\text{Res}_{\bs{z}^*=\bs{0}}\Big[\cphi_I\Big]}{\prod_{i=1}^{n-|I|} (\c{\alpha}_i)} 
        \text{Res}_{\bs{z}^*=\bs{0}}\Big[\res_I\left[\vphi\right]\Big]~,
\end{align}
where, near $\bs{z}^*=\bs{0}$, $\c{\omega}_I\vert_{\bs{z}^*} = \sum_{i=1}^{n-|I|} \c{\alpha}_i\ \d\log z^*_i + \mathcal{O}(\bs{z}^*)$. 
For those familiar with intersection theory, the first term in the product above is a stand-in for the leading term of the maximal primitive for $\cphi_I$.

This formula makes it obvious that the intersection number vanishes if $\cphi_I$ and $\res_I\left[\vphi\right]$ do not share a singularity at any $\bs{z}^* \in \text{Int}_I$. 
For FRW integrals, the $\c{\alpha}_i$ are simply multiples of $\vep$ (intersection points at infinity can introduce non-trivial multiples of $\vep$).

\subsection{Tree-level two-site example \label{app:2site ex}}

As a simple example, we compute the intersection number $\la \cvphi_1 \vert \vth_2 \ra$. 
Using the $\delta_3$ in $\cvphi_1$, we find
\begin{align}
    \la \cvphi_1 \vert \vth_2 \ra 
    &= 
    \bigg\la 
         \vep\
         \d\log \frac{T_1\vert_3}{T_2\vert_3}
    \bigg\vert 
        \text{Res}_3\bigg[ 
            \d\log\frac{T_1}{S_2}
            \wedge\d\log\frac{S_2}{S_3}  
        \bigg]
    \bigg\ra 
    = \vep\
    \bigg\la 
         \d\log \frac{T_1\vert_3}{T_2\vert_3}
    \bigg\vert 
        \d\log \frac{T_1\vert_3}{S_2\vert_3}
    \bigg\ra~.
\end{align}
Supposing that we eliminated $x_2$ when taking the $\delta_3$ residue, there are three twisted intersection points in $x_1$: $\text{Int} = \{ 0, -(X_1 + X_2), \infty\}$. The left and right forms in the intersection number share singularities only at $x_1 = 0$. Thus, the intersection number evaluates to 
\begin{align}
    \la \cvphi_1 \vert \vth_2 \ra 
    =  \vep\ \frac{\text{Res}_{x_1=0}\Big[\d\log \frac{T_1\vert_3}{T_2\vert_3}\Big]}{-\vep} 
        \text{Res}_{x_1=0}\left[\d\log \frac{T_1\vert_3}{S_2\vert_3}\right]
    = -1~,
\end{align}
since $\c{\omega} \sim -\vep\ \d\log z^* + \mathcal{O}(z^*)$ near $z^*=x_1=0$.

\subsection{Tree-level three-site example \label{app:3site ex}}

In this section, we work through a more involved example: the intersection number $\la \cvphi_1 \vert \vphi_1 \ra$ in the context of the tree-level three-site example. Taking care of the $\delta_1$ in $\cvphi_1$, yields 
\begin{align} \label{eq:3site ex}
    \la \cvphi_1 \vert \vphi_1 \ra
    & = \frac{\vep^2}{3}
    \bigg\la    
        \d\log\left(\frac{T_1\vert_1}{T_3\vert_1}\right)
        \wedge
        \d\log\left(\frac{T_3\vert_1}{T_2\vert_1}\right)
    \bigg\vert
        \d\log\left(\frac{T_1\vert_1}{T_3\vert_1}\right)
        \wedge
        \d\log\left(\frac{T_3\vert_1}{T_2\vert_1}\right)
    \bigg\ra 
    \, .
\end{align}
On $X \cap S_1$ there are three finite intersection points: $T_i\vert_1 \cap T_j\vert_1$ for $i,j \in \{1,2,3\}$ and $i \neq j$. Additionally, there are possible intersection points with the hyperplane at infinity $H_\infty$. 
Thus, the set of intersection points are 
\begin{align}
    \text{Int} = \{
        T_1\vert_1 \cap T_2\vert_1,\
        T_1\vert_1 \cap T_3\vert_1,\
        T_2\vert_1 \cap T_3\vert_1,\
        T_2\vert_1 \cap H_\infty,\
        T_1\vert_1 \cap H_\infty,\
        T_3\vert_1 \cap H_\infty
    \}~.
\end{align}
Note that the intersection point $T_3\vert_1 \cap H_\infty$ appears in multiple charts of the $\mathbb{CP}^2$ and should only be counted once.

For the example at hand, only the finite intersection points will contribute since the forms are regular on the hyperplane at infinity. 
Then, for each of these intersection points, we use the corresponding hyperplanes as the normal coordinates $\bs{z}^* = \{T_i\vert_1,T_j\vert_1\}$. In these coordinates, the dual connection becomes $\com\vert_1 = -\vep\ \d\log z_1^* -\vep\ \d\log z_2^* + \mathcal{O}(1)$ near each of the finite intersection points. Using \eqref{eq:CombInt}, the intersection \eqref{eq:3site ex} becomes 
\begin{align}
    \la \cvphi_1 \vert \vphi_1 \ra
    = \frac{\vep^3}{3} \left(\frac{3}{(-\vep)^2}\right)
    = 1~.
\end{align}
The factor of 3 simply comes from the fact that each of the three finite intersection points contribute the same factor of $1/(-\vep)^2$ to the intersection number. 

\section{Explicit presentation of the three-site/five-point differential equation \label{app:3siteDEQ}}

In this appendix, we record the physical differential equation $\mat{A}_\text{phys} = - (\c{\mat{A}}_\text{phys})^\top$. 
Explicitly, the physical alphabet is 
\begin{align} \label{eq:3siteAlpha}
    a_{i=1,\dots,19} = \big\{&
        X_1{+}X_2{+}X_3,\
        X_1{-}Y_1,\
        X_2{+}X_3{-}Y_1,\
        X_1{+}Y_1,\
        X_2{+}X_3{+}Y_1,\
        X_1{-}Y_1{-}2 Y_2, \
        \nn\\&
        X_1{+}X_2{-}Y_2, \
        X_3{-}Y_2, \
        X_3{-}2 Y_1{-}Y_2, \
        X_2{-}Y_1{-}Y_2, \
        X_1{+}X_3{-}Y_1{-}Y_2, \
        \nn\\&
        X_2{+}Y_1{-}Y_2, \
        X_3{+}2 Y_1{-}Y_2, \
        X_1{+}X_2{+}Y_2, \
        X_3{+}Y_2, \
        X_2{-}Y_1{+}Y_2, \
        \nn\\&
        X_1{-}X_3{-}Y_1{+}Y_2, \
        X_2{+}Y_1{+}Y_2, \
        X_1{-}Y_1{+}2 Y_2 
    \big\}~. 
\end{align}
Though a slight abuse of notation, we denote $\d\log(a_i)$ as $a_i$ below in order to fit $\mat{A}_\text{phys}$ on the page
\begin{align} \label{eq:3siteAphys}
\rotatebox{90}{\scalebox{.55}{$
\mat{A}_\text{phys} =    
\left(
\begin{array}{ccccccccccccc}
 3 a_1 & 0 & 0 & 0 & 0 & 0 & 0 & 0 & 0 & 0 & 0 & 0 & 0 \\
 a_3{-}a_4 & 2 a_3{+}a_4 & 0 & 0 & 0 & 0 & 0 & 0 & 0 & 0 & 0 & 0 & 0 \\
 a_7{-}a_{15} & 0 & 2 a_7{+}a_{15} & 0 & 0 & 0 & 0 & 0 & 0 & 0 & 0 & 0 & 0 \\
 a_{18}{-}a_{11} & 0 & 0 & 2 a_{11}{+}a_{18} & 0 & 0 & 0 & 0 & 0 & 0 & 0 & 0 & 0 \\
 a_{14}{-}a_8 & 0 & 0 & 0 & a_8{+}2 a_{14} & 0 & 0 & 0 & 0 & 0 & 0 & 0 & 0 \\
 a_5{-}a_2 & 0 & 0 & 0 & 0 & a_2{+}2 a_5 & 0 & 0 & 0 & 0 & 0 & 0 & 0 \\
 0 & 0 & 0 & 0 & 0 & 0 & a_4{+}2 a_5 & 0 & 0 & 0 & 0 & 0 & 0 \\
 0 & 0 & 0 & 0 & 0 & 0 & 0 & 2 a_{14}{+}a_{15} & 0 & 0 & 0 & 0 & 0 \\
 0 & 0 & 0 & 0 & 0 & 0 & 0 & 0 & \frac{a_5}{2}{+}2 a_{14}{+}\frac{a_{17}}{2} & \frac{a_5}{2}{-}\frac{a_{17}}{2} & 0 & 0 & 0 \\
 0 & 0 & 0 & 0 & 0 & 0 & 0 & 0 & \frac{3 a_5}{2}{-}\frac{3 a_{17}}{2} & \frac{3 a_5}{2}{+}\frac{3 a_{17}}{2} & 0 & 0 & 0 \\
 0 & a_{10}{-}a_{15} & a_{10}{-}a_4 & 0 & 0 & 0 & 0 & 0 & 0 & 0 & a_4{+}a_{10}{+}a_{15} & 0 & 0 \\
 0 & a_{18}{-}a_9 & 0 & a_9{-}a_4 & 0 & 0 & 0 & 0 & 0 & 0 & 0 & a_4{+}a_9{+}a_{18} & 0 \\
 0 & a_{16}{-}a_8 & 0 & 0 & a_{16}{-}a_4 & 0 & 0 & 0 & 0 & 0 & 0 & 0 & a_4{+}a_8{+}a_{16} \\
 0 & 0 & a_6{-}a_{18} & a_{15}{-}a_6 & 0 & 0 & 0 & 0 & 0 & 0 & 0 & 0 & 0 \\
 0 & 0 & a_2{-}a_{12} & 0 & 0 & a_{15}{-}a_{12} & 0 & 0 & 0 & 0 & 0 & 0 & 0 \\
 0 & 0 & 0 & 2 a_2{-}2 a_8 & a_{18}{-}a_2 & a_8{-}a_{18} & 0 & 0 & 0 & 0 & 0 & 0 & 0 \\
 0 & 0 & 0 & 0 & a_2{-}a_{18} & a_8{-}a_{18} & 0 & 0 & a_2{+}\frac{a_{17}}{2}{-}a_{18}{-}\frac{a_5}{2} & {-}\frac{a_5}{2}{+}a_8{-}\frac{a_{17}}{2} & 0 & 0 & 0 \\
 0 & 0 & 0 & 0 & 0 & 0 & 0 & 0 & 0 & 0 & 0 & 0 & 0 \\
 0 & 0 & 0 & 0 & 0 & 0 & 0 & a_4{-}a_{16} & 0 & 0 & 0 & 0 & 0 \\
 0 & 0 & 0 & 0 & 0 & 0 & a_{15}{-}a_{12} & 0 & 0 & 0 & 0 & 0 & 0 \\
 0 & 0 & 0 & 0 & 0 & 0 & a_8{-}a_{18} & 0 & 0 & 0 & 0 & 0 & 0 \\
 0 & 0 & 0 & 0 & 0 & 0 & a_{13}{-}a_{16} & 0 & a_4{-}a_{16}{+}\frac{a_{17}}{2}{-}\frac{a_5}{2} & {-}\frac{a_5}{2}{+}a_{13}{-}\frac{a_{17}}{2} & 0 & 0 & 0 \\
 0 & 0 & 0 & 0 & 0 & 0 & 0 & a_{18}{-}a_2 & 0 & 0 & 0 & 0 & 0 \\
 0 & 0 & 0 & 0 & 0 & 0 & 0 & a_{19}{-}a_{12} & {-}\frac{a_5}{2}{-}a_{12}{+}\frac{a_{17}}{2}{+}a_{19} & {-}\frac{a_5}{2}{+}a_{15}{-}\frac{a_{17}}{2} & 0 & 0 & 0 \\
 0 & 0 & 0 & 0 & 0 & 0 & 0 & 0 & a_2{+}\frac{a_{17}}{2}{-}a_{18}{-}\frac{a_5}{2} & {-}\frac{a_5}{2}{+}a_8{-}\frac{a_{17}}{2} & 0 & 0 & 0 \\
\end{array}
\right.
$}}
\rotatebox{90}{
\scalebox{.55}{$
\qquad
\left.
\begin{array}{cccccccccccc}
 0 & 0 & 0 & 0 & 0 & 0 & 0 & 0 & 0 & 0 & 0 & 0 \\
 0 & 0 & 0 & 0 & 0 & 0 & 0 & 0 & 0 & 0 & 0 & 0 \\
 0 & 0 & 0 & 0 & 0 & 0 & 0 & 0 & 0 & 0 & 0 & 0 \\
 0 & 0 & 0 & 0 & 0 & 0 & 0 & 0 & 0 & 0 & 0 & 0 \\
 0 & 0 & 0 & 0 & 0 & 0 & 0 & 0 & 0 & 0 & 0 & 0 \\
 0 & 0 & 0 & 0 & 0 & 0 & 0 & 0 & 0 & 0 & 0 & 0 \\
 0 & 0 & 0 & 0 & 0 & 0 & 0 & 0 & 0 & 0 & 0 & 0 \\
 0 & 0 & 0 & 0 & 0 & 0 & 0 & 0 & 0 & 0 & 0 & 0 \\
 0 & 0 & 0 & 0 & 0 & 0 & 0 & 0 & 0 & 0 & 0 & 0 \\
 0 & 0 & 0 & 0 & 0 & 0 & 0 & 0 & 0 & 0 & 0 & 0 \\
 0 & 0 & 0 & 0 & 0 & 0 & 0 & 0 & 0 & 0 & 0 & 0 \\
 0 & 0 & 0 & 0 & 0 & 0 & 0 & 0 & 0 & 0 & 0 & 0 \\
 0 & 0 & 0 & 0 & 0 & 0 & 0 & 0 & 0 & 0 & 0 & 0 \\
 a_6{+}a_{15}{+}a_{18} & 0 & 0 & 0 & 0 & 0 & 0 & 0 & 0 & 0 & 0 & 0 \\
 0 & a_2{+}a_{12}{+}a_{15} & 0 & 0 & 0 & 0 & 0 & 0 & 0 & 0 & 0 & 0 \\
 0 & 0 & a_2{+}a_8{+}a_{18} & 0 & 0 & 0 & 0 & 0 & 0 & 0 & 0 & 0 \\
 0 & 0 & 0 & a_2{+}a_8{+}a_{18} & 0 & 0 & 0 & 0 & 0 & 0 & 0 & 0 \\
 0 & 0 & 0 & 0 & a_4{+}a_{15}{+}a_{18} & 0 & 0 & 0 & 0 & 0 & 0 & 0 \\
 0 & 0 & 0 & 0 & 0 & a_4{+}a_{15}{+}a_{16} & 0 & 0 & 0 & 0 & 0 & 0 \\
 0 & 0 & 0 & 0 & 0 & 0 & a_4{+}a_{12}{+}a_{15} & 0 & 0 & 0 & 0 & 0 \\
 0 & 0 & 0 & 0 & 0 & 0 & 0 & a_4{+}a_8{+}a_{18} & 0 & 0 & 0 & 0 \\
 0 & 0 & 0 & 0 & 0 & 0 & 0 & 0 & a_4{+}a_{13}{+}a_{16} & 0 & 0 & 0 \\
 0 & 0 & 0 & 0 & 0 & 0 & 0 & 0 & 0 & a_2{+}a_{15}{+}a_{18} & 0 & 0 \\
 0 & 0 & 0 & 0 & 0 & 0 & 0 & 0 & 0 & 0 & a_{12}{+}a_{15}{+}a_{19} & 0 \\
 0 & 0 & 0 & 0 & 0 & 0 & 0 & 0 & 0 & 0 & 0 & a_2{+}a_8{+}a_{18} \\
\end{array}
\right)
$}}
\end{align}

\bibliographystyle{JHEP}
\bibliography{refs}


\end{document}